\documentstyle[11pt]{article}
\newcommand{\reell}{{\kern+.25em\sf{R}\kern-.78em\sf{I}\kern+.78em\kern-.25em}}
\renewcommand{\theequation}{\arabic{section}.\arabic{equation}}
\begin{document}
\title{Applications of a new proposal for solving
      the ``problem of time'' to some simple
      quantum cosmological models}
\author{Atsushi Higuchi\\
Institut f\"ur theoretische Physik, Universit\"at Bern\\
         Sidlerstrasse 5, CH-3012 Bern, Switzerland\\ and\\
Robert M. Wald\\
Enrico Fermi Institute and Department of Physics, University of
         Chicago\\  5640 S. Ellis Ave., Chicago, IL 60637-1433}
\date{July 26, 1994}
\maketitle
\begin{abstract}
We apply a recent proposal for defining states and
observables in quantum gravity to some simple models. First, we
consider the toy model of a Klein-Gordon particle in an external potential in
Minkowski spacetime and compare our proposal to the theory obtained by
deparametrizing with respect to a choice of
time slicing prior to quantization. We
show explicitly that the dynamics defined by the deparametrization
approach depends upon the choice of
time slicing. On the other hand, our proposal automatically yields a well
defined dynamics -- manifestly independent of the choice of time slicing at
intermediate times -- but there is a ``memory effect": After the potential
is turned off, the dynamics no longer returns to the standard, free
particle dynamics. Next, we apply our proposal to the closed
Robertson-Walker quantum cosmology with a homogeneous massless scalar field.
We choose our time variable to be the size of the universe,
so the only dynamical variable is the scalar field.
It is shown that the resulting theory has the expected semi-classical
behavior up to the classical
turning point from expansion to contraction, i.e., given a classical solution
which expands for much longer than the Planck time, there is a quantum
state whose dynamical evolution closely approximates this classical
solution during the expanding phase.
However, when the ``time'' takes a
value larger than the classical maximum, the scalar field becomes ``frozen"
at the value it had when it entered the classically forbidden region.
The Taub model, with and without a homogeneous scalar field,
also is studied, and similar behavior is found. In an Appendix, we derive
the form of the
Wheeler-DeWitt equation on mini\-superspace for the Bianchi
models by performing a proper quantum
reduction of the momentum constraints; this equation differs from the
usual form of the Wheeler-DeWitt equation, which is
obtained by solving the momentum constraints classically, prior to
quantization.
\end{abstract}

\begin{flushleft}
{\bf PACS \#:} 04.60.+n, 03.70.+k, 04.20.Cv
\end{flushleft}
\oddsidemargin=0in
\evensidemargin=0in
\topmargin=-.5in
\textwidth=6.7in
\textheight=9in
\hsize=6.27truein
\newpage
\begin{section}{Introduction}
\label{intro}
The ``problem of time'' refers to the difficulties in defining a Hilbert space
of states, observables, and a nontrivial notion of dynamics in a theory
where the spacetime metric is itself a dynamical variable, so no
background
metrical or causal structure is present. In the canonical
approach to quantum gravity, this difficulty manifests itself
directly by the fact that ordinary time evolution corresponds to a
diffeomorphism (i.e., a
gauge transformation), and the Hamiltonian is thereby constrained to
vanish \cite{COM1}.
As a result, the ``wavefunction of the universe"
in quantum gravity does not depend on ``time".
This necessitates a serious reconsideration of
the structure and interpretation
of the theory provided by a naive application of the usual canonical
quantization prescription.

Recently one us proposed \cite{W} a prescription for defining states,
observables, and dynamics in quantum gravity by making
use of the background
metrical and causal structure which is available in superspace.
The main purpose
of this paper is to further elucidate this proposal by applying it to some
simple models.
In Sec.\ \ref{Prop} we briefly outline the proposal in the context of
quantum cosmology.
In Sec.\ \ref{Adm} we study a toy model of massless scalar particle
in an external potential in
Minkowski spacetime -- which shares some of the key features of the
quantum-cosmological models -- for the purpose of comparing the proposal with
the ``reduced theory" obtained by deparametrization with respect to a choice of
time variable.  We show explicitly in this model
that the dynamics defined by the deparametrization
approach depends upon the choice of time variable.
In contrast, the dynamics of our
proposal is well defined, but there is a ``memory effect'',
wherein the dynamics
does not return to that of standard free particle theory
after the potential is turned
off. In Sec.\ \ref{Hc} we then turn to an analysis of some
features of the proposal in the context of
general homogeneous cosmologies with a scalar field.
In Sec.\ \ref{Rw} we apply the proposal to the closed Robertson-Walker
cosmology with a homogeneous scalar field.  We find that the semi-classical
description is a good approximation right up to the classical turning point
from expansion to contraction. We also find that
after this stage, the scalar field -- which
is the only dynamical variable -- is frozen, i.e., it does not evolve with
``time'', which is chosen to be the size of the universe.
In Sec.\ \ref{Taub} we repeat the analysis of Sec.\ \ref{Rw} for the
Taub universe with and without a homogeneous scalar field and find
qualitatively similar results. Finally, in Appendix A, we derive the
Wheeler-DeWitt equation for the Bianchi models by performing a quantum
(rather than classical) reduction of the momentum constraints.
\end{section}
\setcounter{equation}{0}

\begin{section}{The proposal}
\label{Prop}
First we briefly review the proposal of Ref.\ \cite{W} in the context
of quantum cosmology.
The dynamical variables in quantum cosmology are the homogeneous
spatial metric
$h_{ab}$ and possible homogeneous matter fields together with the conjugate
momenta of these variables. Let us denote the configuration variables
(i.e., coordinates on mini\-superspace) by $q^{A}$
($A = 1,\ldots, N$) and their conjugate momenta by $\pi_A$.  Then, as will be
discussed in detail in Sec.\ \ref{Hc}
(see also Appendix A), for the models under
consideration here, the
classical Hamiltonian constraint takes the form
\begin{equation}
H = G^{AB}\pi_A\pi_B + V(q) = 0\,.   \label{2suha}
\end{equation}
Here the supermetric, $G_{AB}$, has Lorentz signature,
$-+\cdots+$, and, in our models, mini\-superspace is globally hyperbolic
in this metric.
Furthermore, $G_{AB}$ possesses a one-parameter group of timelike conformal
isometries corresponding to scaling transformations of
the spatial metric, i.e.,
\begin{equation}
h_{ab} \to e^{2\alpha}  h_{ab}\,.   \label{alpha}
\end{equation}
(This one-parameter group of conformal isometries extends to full
superspace \cite{KAR}, and,
thus, is not an artifact of the simplicity of our models.)
For convenience, we shall
redefine $G_{AB}$, if necessary, by a
conformal transformation so as to make these
conformal isometries be true isometries.

We take the quantum version of the
Hamiltonian constraint -- i.e., the Wheeler-DeWitt equation -- to be
a Klein-Gordon equation of the form
\begin{equation}
\left[ -\nabla^A\nabla_A + \xi R + V(q)\right]\Phi = 0\,.  \label{kgeq}
\end{equation}
Since $G_{AB}$ is naturally defined only up to conformal equivalence
class, it would appear most natural to choose $\xi$ so that (\ref{kgeq})
is conformally invariant \cite{MH} (see, however, the discussion
following Eq.\ (\ref{conc}) of Appendix A).

To define a Hilbert space structure on an appropriate subspace
of solutions, $\Phi$, of this equation, we make use of the natural
 (real) symplectic product, $\Omega$, given by
\begin{equation}
\Omega(\Phi_1,\Phi_2) =
\int_{\Sigma}\left[ \Phi_2\nabla_A\Phi_1 - \Phi_1\nabla_A\Phi_2\right]
d\Sigma^A\,, \label{2inn}
\end{equation}
where $\Sigma$ is any Cauchy surface.
Note that $\Omega$ is conformally invariant \cite{HK} provided that we scale
the wavefunction $\Phi$ as $\Phi \to \lambda(q)^{1-N/2}\Phi$ under
$G_{AB} \to \lambda(q)^2 G_{AB}$.
With our choice of supermetric, minisuperspace
is static as a spacetime under the isometries
defined by Eq.\ (\ref{alpha}), so if $V(q)$ were
$\alpha$-independent, then the solutions that are positive frequency
with respect to $\alpha$ could be
chosen to form the Hilbert space \cite{AMK}, with inner product
\begin{equation}
\langle \Phi_1, \Phi_2 \rangle_{KG} = -i\Omega(\bar{\Phi}_1,\Phi_2)\,.
\label{inner}
\end{equation}
However, the potential $V(q)$
depends on $\alpha$ and, in fact, there is no symmetry of the
super-Hamiltonian,
$H$, that can be used to define positive frequency solutions \cite{KAR}.
Nevertheless, we have $V \to 0$ as $\alpha \to -\infty$. Thus, the
Wheeler-DeWitt equation (\ref{kgeq}) possesses an asymptotic symmetry at
``early times".
It was conjectured in Ref.\ \cite{W} that -- at least in a suitable class of
homogeneous cosmologies -- this asymptotic symmetry enables
one to construct the Hilbert space, ${\cal H}$,
of solutions that are positive frequency with respect to $\alpha$ in the
limit $\alpha \to -\infty$, with inner product (\ref{inner}).
We shall discuss this issue further in Sec.\ \ref{Hc}.

Assuming that the Hilbert space, ${\cal H}$, is obtained as we
described,
we complete our construction of a quantum theory corresponding to the
classical super-Hamiltonian (\ref{2suha}) by specifying the
self-adjoint operators on ${\cal H}$ which represent
``position and momentum" observables at a given ``instant of time".
We take the notion corresponding to an ``instant of time" in this theory
to be the specification of a Cauchy hypersurface, $\Sigma$,
in mini\-superspace. There exists
a standard prescription for the construction of the desired position
and momentum operators on the Hilbert space $L^2(\Sigma)$ (see,
e.g., appendix C of Ref.\ \cite{ash}) but
our inner product is not
the $L^2$ inner product but the Klein-Gordon one.
Nevertheless,
any $C^1$ solution to the Klein-Gordon equation which lies
in our one-particle Hilbert space ${\cal H}$
is uniquely determined by its value on
$\Sigma$ \cite{W}, since,
by Eqs.\ (\ref{2inn}) and (\ref{inner}),
the difference between two solutions in ${\cal H}$
having the same restriction to $\Sigma$ must have vanishing Klein-Gordon
norm.
As in Ref.\ \cite{W},
we assume that the subspace, ${\cal D}$,
of $C^1$ solutions in ${\cal H}$ whose restriction to $\Sigma$
lies in $L^2(\Sigma)$ is dense both
as a subspace of ${\cal H}$ and as a subspace of $L^2(\Sigma)$,
and that if a sequence in ${\cal D}$ converges in both
${\cal H}$ and $L^2(\Sigma)$, then its limit in ${\cal H}$ is nonzero if and
only if its limit in $L^2(\Sigma)$ is nonzero.
If we view the Klein-Gordon inner product (\ref{inner})
as a quadratic form defined on a dense domain
in $L^2(\Sigma)$, it then follows that one can write
\begin{equation}
\langle \Phi_1,\Phi_2\rangle_{KG}
= (B(\Phi_1|_{\Sigma}), B(\Phi_2|_{\Sigma}))_{L^2}\,,     \label{defa}
\end{equation}
where $B: L^2(\Sigma) \to L^2(\Sigma)$ is a positive self-adjoint
operator. We then define the position and momentum observables at
``time" $\Sigma$ by
\begin{equation}
\langle \Phi_1\vert {\cal O}\vert \Phi_2\rangle_{KG} \equiv
(B(\Phi_1|_{\Sigma}),  \tilde{{\cal O}}B(\Phi_2|_{\Sigma}))_{L^2}\,,
\end{equation}
where $\tilde{{\cal O}}$ is the standard representation of the observable
on $ L^2(\Sigma)$. Note that the
operator $B$ defined by Eq.\ (\ref{defa}) depends, in general, on the entire
``history" of the mini\-superspace from the ``asymptotic past" (i.e.,
$\alpha \to -\infty$) to ``time" $\Sigma$, so the definition of ${\cal O}$ is
nonlocal in ``time" as well as ``space".

The time evolution of a state $\Phi \in {\cal H}$ is, of course,
given simply by Eq.\ (\ref{kgeq}), but the observables, ${\cal O}$, are
not represented in a simple manner on ${\cal H}$. It is useful, therefore,
to derive the evolution equation for the corresponding states
$\Psi \equiv B (\Phi|_\Sigma)$
in $ L^2(\Sigma)$, where the observables, $\tilde{{\cal O}}$,
take a simple form.
Let $\Sigma_t$ be a one-parameter family of Cauchy surfaces labeled by a time
function $t$, and let $t^a$ be a time evolution vector field, satisfying
$t^a \nabla_a t = 1$, which may be decomposed into a lapse and shift via
\begin{equation}
t^a = Nn^a + N^a\,,
\end{equation}
where $n^a$ denotes the unit normal to $\Sigma_t$, and where $N^a$ is
tangential to $\Sigma_t$. As noted above, for
any $\Phi \in {\cal H}$
which is $C^1$, the quantity $(n^a \nabla_a \Phi)|_{\Sigma_t}$ is uniquely
determined by $\Phi|_{\Sigma_t}$.
Assuming that $(n^a \nabla_a \Phi)|_{\Sigma_t}$ lies in
$L^2(\Sigma_t)$ for at least a dense (in $L^2(\Sigma_t)$)
subspace of $\Phi \in {\cal D}$, we obtain, for each $t$ an operator
$C_t: L^2(\Sigma_t) \to L^2(\Sigma_t)$ such that
\begin{equation}
(n^a \nabla_a \Phi)|_{\Sigma_t} = -iC_t (\Phi|_{\Sigma_t})\,.
\label{Ct}
\end{equation}
In the following we shall omit writing ``$|_{\Sigma_t}$", i.e., it should
be understood that, for example, $C_t \Phi$ means
$C_t (\Phi|_{\Sigma_t})$.
It follows immediately from Eqs.\ (\ref{2inn}) and (\ref{inner}) that for
$\Phi_1, \Phi_2$
in the subspace of ${\cal D}$ on which $C_t$ is defined,
we have
\begin{eqnarray}
\langle \Phi_1,\Phi_2\rangle_{KG} &=&
i(\Phi_1, n^a \nabla_a \Phi_2)_{L^2(\Sigma_t)}
-i(n^a \nabla_a \Phi_1, \Phi_2)_{L^2(\Sigma_t)} \nonumber \\
&=& (\Phi_1, C_t \Phi_2)_{L^2(\Sigma_t)} +
(C_t \Phi_1, \Phi_2)_{L^2(\Sigma_t)} \nonumber \\
&=& (\Phi_1, (C_t + C^{\dag}_t) \Phi_2)_{L^2(\Sigma_t)}\,.
\end{eqnarray}
By comparison with Eq.\ (\ref{defa}), we obtain
\begin{equation}
B_t = (C_t + C^{\dag}_t)^{1/2}\,.
\label{BC}
\end{equation}
By brute force substitution, we obtain as the evolution equation for
$\Psi_t \equiv B_t \Phi_t$
\begin{eqnarray}
\frac{\partial \Psi_t}{\partial t} &=&
\frac{\partial \ }{\partial t} (B_t \Phi_t) \nonumber \\
&=& \frac{\partial B_t}{\partial t} \Phi_t +
B_t \frac{\partial \Phi_t}{\partial t} \nonumber \\
&=& \frac{\partial B_t}{\partial t} B^{-1}_t \Psi_t
- i B_t NC_t B^{-1}_t \Psi_t
+ B_t N^a \nabla_a (B^{-1}_t \Psi_t)  \nonumber \\
&=& -iH_{\it eff} \Psi_t\,,
\end{eqnarray}
where the effective Hamiltonian for evolution in $L^2(\Sigma_t)$
is given by
\begin{equation}
H_{\it eff} \equiv i\frac{\partial B_t}{\partial t} B^{-1}_t
+ B_t NC_t B^{-1}_t + iB_t N^a \nabla_a B^{-1}_t\,.
\label{Heff}
\end{equation}
Since $B_t$ is given in terms of $C_t$ by Eq.\ (\ref{BC}), the effective
Hamiltonian is determined directly by $C_t$.
Note that since the natural volume element on $\Sigma_t$ will, in
general, be ``time dependent" (i.e., not Lie-derived by $t^a$),
the natural $L^2$ inner product on $\Sigma_t$ will similarly
be time dependent, and
$H_{\it eff}$ will not be self-adjoint. However, we could define a
self-adjoint evolution by working with a fixed ``coordinate" volume
element on $\Sigma_t$ to define a fixed inner product for $L^2(\Sigma_t)$
or, equivalently, working with densitized versions of
$\Psi_t$. We have chosen not to do so here since this would
further complicate the formulas, and, in all of the applications in this
paper, the natural volume element on $\Sigma_t$ will be Lie
derived by our time evolution vector field $t^a$.

The above general relations simplify considerably for the quantum
cosmological models we shall consider in this paper. First, we shall
choose as our Cauchy surfaces which foliate mini\-superspace the
hypersurfaces orthogonal to
a timelike Killing field $t^a$.
We choose $t^a$ as the time evolution
vector field, so we have $N^a = 0$ and the
Cauchy surfaces are labeled by the time
function $t$ such that $t^a\nabla_a t = 1$.
Furthermore, in our models, we have
$N = 1$. Most importantly, in our models the
$t$-dependence is separable, so
we may work with (un-normalizable) basis functions
for ${\cal H}$ of the form
\begin{equation}
\Phi_{\omega\sigma}(t,q^i) = \frac{f_{\omega}(t)}{\sqrt{2\omega}}
S_{\omega\sigma}(q^i)\,,  \label{2sep}
\end{equation}
where  $f_{\omega}(t)$ is such that
$f_{\omega}(t) \to e^{-i\omega t}$ as $t \to -\infty$.
Here $q^i$ are the
configuration variables apart from $t$, and
the label $\sigma$ represents the quantum numbers other than $\omega$.
The normalization of
$S_{\omega\sigma}(q^i)$ is chosen to be
$(S_{\omega'\sigma'},S_{\omega\sigma})_{L^2}
= \delta(\omega'-\omega)\delta(\sigma',\sigma)$, where
$\delta(\sigma',\sigma)$ is an appropriate delta-function for the label
$\sigma$, so that
$\langle\Phi_{\omega'\sigma'},\Phi_{\omega\sigma}\rangle_{KG} =
\delta(\omega'-\omega)\delta(\sigma',\sigma)$. Note that the
conservation of the Klein-Gordon inner product together
with the asymptotic form of $f_{\omega}(t)$
as $t \to -\infty$ implies that
\begin{equation}
\bar{f}_{\omega}(t)f'_{\omega}(t)
- f_{\omega}(t)\bar{f}'_{\omega}(t)
= -2i\omega\,,  \label{2wron}
\end{equation}
where $f' \equiv df/dt$.

Differentiating Eq.\ (\ref{2sep}) with respect to $t$, we obtain
\begin{eqnarray}
\frac{\partial \Phi_{\omega\sigma}}{\partial t}
&=& \frac{f'_{\omega}(t)}{\sqrt{2\omega}}
S_{\omega\sigma}(q^i) \nonumber \\
&=& \frac{f'_{\omega}(t)}{f_{\omega}(t)} \Phi_{\omega\sigma}\,.
\end{eqnarray}
Thus, by inspection, the operator $C_t$ defined by Eq.\ (\ref{Ct}) is
diagonal in this basis, and we have
\begin{equation}
C_t\Phi_{\omega\sigma}  =
i\frac{f'_{\omega}(t)}{f_{\omega}(t)}
\Phi_{\omega\sigma}\,.
\end{equation}
Using Eq.\ (\ref{2wron}), we obtain
\begin{eqnarray}
B_t\Phi_{\omega\sigma} & = &
(C_t+ C^{\dag}_t)^{1/2} \Phi_{\omega\sigma} \nonumber \\
& = & \frac{\sqrt{2\omega}}{|f_{\omega}(t)|} \Phi_{\omega\sigma} \,.
\end{eqnarray}
Substitution of the above expressions for $C_t$ and $B_t$
together with $N=1$ and $N^a = 0$ into Eq.\ (\ref{Heff}) then yields
\begin{eqnarray}
H_{\it eff} \Psi_{\omega\sigma} & = &
\left[ i \frac{d}{dt}\left(\frac{1}{|f_{\omega}(t)|}\right)|f_{\omega}(t)| +
i\frac{f'_{\omega}(t)}{f_{\omega}(t)}\right]
\Psi_{\omega\sigma} \nonumber \\
& = & \frac{\omega}{|f_{\omega}(t)|^2} \Psi_{\omega\sigma}\,,
\label{heffqc}
\end{eqnarray}
where Eq.\ (\ref{2wron}) was used again in the last line.
\end{section}
\setcounter{equation}{0}

\begin{section}{Comparison with Deparametrization}
\label{Adm}
An alternative scheme to the one presented in the previous section for
constructing a quantum theory corresponding to a classical theory with
super-Hamiltonian of the form (\ref{2suha}) is to ``deparametrize" the
theory prior to quantization in the manner proposed by Arnowitt,
Deser, and Misner \cite{ADM}. In this procedure, we first choose a time
function, $t$, whose level surfaces are Cauchy surfaces, $\Sigma_t$,
and choose
a time evolution vector field, $t^a$, satisfying $t^a \nabla_a t = 1$. We
then solve Eq.\ (\ref{2suha}) for the momentum, $\pi_t$, canonically
conjugate to $t$, thereby expressing $\pi_t$ as a function of $t$ and the
remaining dynamical variables $(q^{i}$, $\pi_{i})$. The Hilbert space of
states at time $t$ is then taken to be $L^2(\Sigma_t)$, and the quantum
operators corresponding to $q^{i}$ and $\pi_{i}$ at time $t$ are defined by
the standard prescription for $L^2(\Sigma_t)$. Dynamical evolution then is
defined by
\begin{equation}
\frac{d\Phi}{dt} = -i H_{\it ADM} \Phi \,,
\end{equation}
where the ADM Hamiltonian is given by
\begin{equation}
H_{\it ADM} \equiv -\pi_t(t, q^i,\pi_i)\,. \label{hadm}
\end{equation}
Classically, this Hamiltonian generates the dynamics of
$\left\{ q^{i}, \pi_{i} \right\}$, with respect to time $t$, as can
be verified by using the
original Hamilton's equations with the super-Hamiltonian $H$.

Since the proposal of the previous section also can be viewed as a theory
defined on the Hilbert spaces $L^2(\Sigma_t)$ with the
standard definitions
of position and momentum operators, it differs from the quantum theory
obtained by the above
deparametrization method only via the use of the Hamiltonian (\ref{Heff})
rather than (\ref{hadm}). In this section, we shall compare the two
approaches for the ``toy model" of a
relativistic particle in four dimensions with a
potential of compact spacetime support, whose super-Hamiltonian
is given by
\begin{equation}
H= -p_{0}^2 + {\bf p}^2 + \lambda V(x_0,{\bf x})  = 0\,. \label{clham}
\end{equation}
We study this model to first order in a perturbation
series in $\lambda$. We shall show explicitly in this model that
-- as expected (see, e.g., \cite{KUCH}) -- the
dynamical evolution defined by the deparametrization procedure
depends upon the choice of time variable, $t$, used to perform the
deparametrization. The quantum theory defined by the proposal of
the previous section is manifestly independent of a choice of
slicing, but we shall show explicitly the presence of a ``memory effect",
which illustrates the nonlocal-in-time character of dynamical evolution.

First we solve the Klein-Gordon equation (\ref{kgeq}) for this model to
lowest order in $\lambda$. We have
\begin{equation}
\left( -\frac{\partial^2\ \ }{\partial x_0^2} +
\frac{\partial^2\ }{\partial{\bf x}^2}\right)\Phi = \lambda
V(x_0,{\bf x})
\Phi\,.  \label{deq}
\end{equation}
Consider the ``unperturbed" wavefunction
\begin{equation}
\Phi^{(0)} = \int \frac{d^3{\bf p}}{(2\pi)^3 2p} f({\bf p})
e^{-ipx_0+i{\bf p}\cdot{\bf x}} \,,
\label{psi0}
\end{equation}
where $p \equiv |{\bf p}|$, and write the solution to (\ref{deq})
to first order in $\lambda$ as
\begin{equation}
\Phi = \Phi^{(0)} + \Phi^{(1)}
\end{equation}
with the boundary condition
$\Phi^{(1)} \rightarrow 0$ for $x_{0} \rightarrow -\infty$.
Note that the Klein-Gordon norm of $\Phi^{(0)}$ is
\begin{equation}
\langle\Phi^{(0)}, \Phi^{(0)}\rangle_{KG}
= \int \frac{d^3{\bf p}}{(2\pi)^3 2p}
\left| f({\bf p}) \right|^2\,.
\end{equation}

The first order in $\lambda$ correction, $\Phi^{(1)}$,
to $\Phi$ due to the presence of the
potential is given by
\begin{equation}
\Phi^{(1)}(x_0, {\bf x})
= \lambda\int dx_0' d^3{\bf x}'
G_{KG}(x_0,{\bf x};x_0',{\bf x}')V(x_0',{\bf x}')
\Phi^{(0)}(x_0^{\prime}, {\bf x}')\,,
\end{equation}
where $G_{KG}$ denotes
the retarded Green function for the Klein-Gordon operator,
\begin{equation}
G_{KG}(x_0,{\bf x};x_0',{\bf x}')  =  -i\theta(x_0-x_0')
\int
\frac{d^3{\bf p}'}{(2\pi)^3 2 p'}
 \left[
e^{-ip'(x_0-x_0')+i{\bf p}'\cdot({\bf x}-{\bf x}')}
- e^{ip'(x_0-x_0')+i{\bf p}'\cdot({\bf x}-{\bf x}')}\right]\,.
\end{equation}
Let $x_0^{(M)}$ denote the maximum value of $x_0$ such that
$V(x_0, {\bf x}) \neq 0$ for some ${\bf x}$.
Then we have for $x_0 > x_0^{(M)}$,
\begin{eqnarray}
\Phi^{(1)}(x_0, {\bf x})
 & = & \lambda\int dx_0' d^3{\bf x}'
G_{KG}(x_0,{\bf x};x_0',{\bf x}')V(x_0',{\bf x}')
\Phi^{(0)} (x_0^{\prime}, {\bf x}') \nonumber \\
& = & -i \lambda\int
\int\frac{d^3{\bf p}}{(2\pi)^3 2p}
\frac{d^3{\bf p}'}{(2\pi)^3 2p'}
\left[ \tilde{V}(p' - p,{\bf p}'-{\bf p})f({\bf p})
e^{-ip'x_0+i{\bf p}'\cdot {\bf x}}\right. \nonumber \\
 & & \ \ \ \ \ \ \ \ \ \ \ \ \ \ \ \ \
\left. -\tilde{V}(-p' - p,{\bf p}'-{\bf p})f({\bf p})
e^{ip'x_0 + i{\bf p}'\cdot {\bf x}}\right]\,,  \label{D}
\end{eqnarray}
where
\begin{equation}
\tilde{V}(p_0,{\bf p}) \equiv \int dx_0 d^3{\bf x}
V(x_0,{\bf x})e^{ip_0 x_0 - i{\bf p}\cdot{\bf x}}\,.
\end{equation}

Consider now the quantum theory obtained by deparametrizing with respect
to the time variable $t = x_0$.
Then we have
\begin{equation}
\pi_t^2 = p^2 + \lambda V (t, {\bf x})\,.
\label{pit}
\end{equation}
The first point to note is that if $V$ becomes negative, a real solution,
$\pi_t$, of Eq.\ (\ref{pit}) will not exist for all values of
$t , ~ {\bf x}$ and
${\bf p}$, and the quantization prescription will break down.
In order to avoid this difficulty, we restrict consideration to potentials
which are everywhere non\--negative.
(Note that, in contrast, the proposal given
in the previous section does not require any such restriction on $V$.)
If $V$ is non\--negative, we have
\begin{equation}
H_{\it ADM} = - \pi_t = \sqrt{p^2 + \lambda V (t, {\bf x})} \,.
\end{equation}
One of the nice features of this model is that the definition of $H_{\it ADM}$
as a quantum operator is unambiguous:  no ``factor ordering'' ambiguities
occur for defining the operator $p^2 + \lambda V$, and, by continuity, one
must choose the positive square root of the operator in order to get
agreement with the standard free particle theory at $\lambda = 0$.
Thus, the dynamical evolution equation in this approach is
\begin{equation}
i\frac{\partial\ }{\partial x_0}\Psi_{\it ADM}
= \left[ -\frac{\partial^2\ }{\partial {\bf x}^2} + \lambda
V(x_0,{\bf x})\right]^{1/2}
\Psi_{\it ADM}\,.  \label{eqadm}
\end{equation}

Now consider the solution to Eq.\ (\ref{eqadm})
to first order in $\lambda$
of the form
\begin{equation}
\Psi_{\it ADM} = \Psi_{\it ADM}^{(0)} + \Psi^{(1)}_{\it ADM}
\end{equation}
with $\Psi^{(1)}_{\it ADM} \rightarrow 0$ for $x_{0} \rightarrow -\infty$,
where
\begin{equation}
\Psi_{\it ADM}^{(0)} = \int \frac{d^3{\bf p}}{(2\pi)^3 \sqrt{2p}} f({\bf p})
e^{-ip x_0+i{\bf p}\cdot{\bf x}}\,.  \label{psadm}
\end{equation}
Note that $\Psi_{\it ADM}^{(0)}$ corresponds precisely to
$\Phi^{(0)}$, Eq.\ (\ref{psi0}); the difference in normalization arises from
the fact that $\Psi_{\it ADM}^{(0)}$ is normalized via the $L^2$ inner product,
whereas $\Phi^{(0)}$ was normalized via the Klein-Gordon inner product.

The square root operator in (\ref{eqadm}) acts on $e^{i{\bf p}\cdot{\bf x}}$
to order $\lambda$ as
\begin{equation}
 \left[ -\frac{\partial^2\ }{\partial {\bf x}^2} + \lambda
V(x_0,{\bf x})\right]^{1/2}
e^{i{\bf p}\cdot{\bf x}}
\approx
 pe^{i{\bf p}\cdot{\bf x}} +
\lambda
\int\frac{d^3{\bf p}'}{(2\pi)^3}\frac{\tilde{V}_s(x_0,{\bf p}'-{\bf p})}
{p+p'}e^{i{\bf p}'\cdot{\bf x}} \,,
\label{sqrtop}
\end{equation}
where
\begin{equation}
\tilde{V}_s(x_0,{\bf p}) \equiv \int d^3{\bf x} V(x_0,{\bf x})
e^{-i{\bf p}\cdot{\bf x}}\,.
\end{equation}
One can readily verify that the square of this
operator is indeed
$-\partial^2/\partial {\bf x}^2 + \lambda V(x_0,{\bf x}) + O(\lambda^2)$.
{}From Eqs.\ (\ref{eqadm}) and (\ref{sqrtop}), we find that the first order
correction, ${\Psi}_{\it ADM}^{(1)}$,
to $\Psi_{\it ADM}$ due to the presence of $V$ satisfies
\begin{equation}
\left( i\frac{\partial\ }{\partial x_0} - \left|
-\frac{\partial^2 \ }{\partial {\bf x}^2}\right|^{1/2}\right)
\Psi_{\it ADM}^{(1)}
= \lambda\int \frac{d^3{\bf p}}{(2\pi)^3\sqrt{2p}}
\frac{\tilde{V}_s(x_0,{\bf p}'-{\bf p})}{p + p'}
f({\bf p})e^{i{\bf p}'\cdot{\bf x}}\,.
   \label{A}
\end{equation}
Using the retarded Green function for the Schr\"odinger operator
appearing on the left side of this equation
\begin{equation}
G_{\it ADM}(x_0,{\bf x};x_0',{\bf x}')
 = -i\theta(x_0-x_0')\int\frac{d^3{\bf p}}{(2\pi)^3}
e^{-ip(x_0-x_0')+ i{\bf p}\cdot({\bf x}-{\bf x}')}\,,
\end{equation}
we find for $x_0 > x_0^{(M)}$
\begin{equation}
\Psi_{\it ADM}^{(1)}  =
-i\lambda \int
\frac{d^3{\bf p}}{(2\pi)^3\sqrt{2p}}
\frac{d^3{\bf p}'}{(2\pi)^3}
\frac{\tilde{V}(p'- p,{\bf p}'-{\bf p})}
{p+p'}f({\bf p})
e^{-ip'x_0+ i{\bf p}'\cdot{\bf x}}\,.
   \label{C}
\end{equation}

In the region $x_0 > x_0^{(M)}$,
we can associate to
$\Psi_{\it ADM}$
a positive frequency solution $\Phi_{\it ADM}$ to the free Klein\--Gordon
equation by the correspondence
\begin{equation}
\Phi_{\it ADM} = (-4 \nabla^2)^{-1/4} \Psi_{\it ADM} \,.
\end{equation}
Writing $\Phi_{\it ADM} = \Phi_{\it ADM}^{(0)} + \lambda \Phi_{\it ADM}^{(1)}+
O(\lambda^2)$, we have
\begin{eqnarray}
\Phi_{\it ADM}^{(0)} & = & \Phi^{(0)}\,, \\
\Phi_{\it ADM}^{(1)} & = &
-i\lambda \int
\frac{d^3{\bf p}}{(2\pi)^3\sqrt{2p}}
\frac{d^3{\bf p}'}{(2\pi)^3\sqrt{2p'}}
\frac{\tilde{V}(p'-p,{\bf p}'-{\bf p})}
{p+p'}f({\bf p})
e^{-ip'x_0+ i{\bf p}'\cdot{\bf x}}\,,
\label{phiadm1}
\end{eqnarray}
where $\Phi^{(0)}$ was given by Eq.\ (\ref{psi0}).
Note that $\Phi_{\it ADM}^{(1)}$ differs
from $\Phi^{(1)}$ in Eq.\ (\ref{D}).
Indeed, unlike $\Phi^{(1)}_{\it ADM}$,
$\Phi^{(1)}$ is not even a purely positive frequency solution.
However, we can consider a WKB expansion of $\Phi_{\it ADM}^{(1)}$
and $\Phi^{(1)}$.
For $V$ smooth, the negative frequency part of $\Phi^{(1)}$ is
``nonperturbative'' in such an expansion, i.e., it does not appear to any
finite order in $\hbar$.
Furthermore, we have verified that to first order in $\hbar , ~
\Phi_{\it ADM}^{(1)}$ and $\Phi^{(1)}$ agree.
This result is consistent with Barvinsky's \cite{BAR} much more
general arguments
in the context of quantum field theory that
``reduced phase space quantization'' agrees
with ``Dirac quantization'' at one\--loop order.
However, in our model,
we find that $\Phi_{\it ADM}^{(1)}$ and $\Phi^{(1)}$ differ
at order $\hbar^2$.

Now we consider a second
deparametrization scheme obtained by using a boosted Lorentz
frame, i.e., we choose
$t = x_{0}^{\beta} \equiv x_0 \cosh\beta + x_1 \sinh\beta$.
The initial ADM wavefunction which corresponds to
$\Psi_{\it ADM}^{(0)}$ given by (\ref{psadm}) is
\begin{equation}
\Psi_{\it ADM}^{(0)\beta} = \int \frac{d^3{\bf p}}{(2\pi)^3 \sqrt{2p}}
f({\bf p}_{-\beta})
e^{-ip x_0^{\beta}+i{\bf p}\cdot{\bf x}^{\beta}}\,,
\end{equation}
where ${\bf x}^{\beta}
\equiv (x_1\cosh\beta + x_0\sinh\beta, x_2,x_3)$ and
${\bf p}_{-\beta}
\equiv (p_1\cosh\beta - p_0\sinh\beta, p_2, p_3)$\,.
If we start from this
initial wavefunction, we obtain for the lowest order correction,
$\Psi_{\it ADM}^{(1)\beta}$, an expression of the
form (\ref{C}) with $x_0$ replaced by
$x_0^{\beta}=x_0\cosh\beta + x_1\sinh\beta$, $x_1$ by
$x_1^{\beta} = x_1\cosh\beta + x_0\sinh\beta$,
$f({\bf p})$ by $f({\bf p}_{-\beta})$ and
$\tilde{V}(p'-p,{\bf p}' - {\bf p})$ by
$\tilde{V}(p_{-\beta}'-p_{-\beta},{\bf p}_{-\beta}' - {\bf p}_{-\beta})$.
The corresponding solution, $\Phi_{\it ADM}^{(1)\beta}$,
of the free Klein-Gordon equation at late times is given by
\begin{equation}
\Phi_{\it ADM}^{(1)^{\beta}} =
-i\lambda \int
\frac{d^3{\bf p}}{(2\pi)^3 2p}
\frac{d^3{\bf p}'}{(2\pi)^3 2p'}
\frac{2\sqrt{p_{\beta}p'_{\beta}}\tilde{V}(p'-p,{\bf p}'-{\bf p})}
{p_{\beta}+p'_{\beta}}f({\bf p})
e^{-ip'x_0+ i{\bf p}'\cdot{\bf x}}\,,
\end{equation}
where ${\bf p}_{\beta} = (p_1\cosh\beta + p_0\sinh\beta, p_2,p_3)$ and
$p_{\beta} = |{\bf p}_{\beta}|$.
This disagrees with the solution $\Phi_{\it ADM}^{(1)}$, Eq.\ (\ref{phiadm1}),
obtained with the original choice of time slicing.
Hence, if we physically identify ADM wavefunctions on different slices
in the region to the future of the potential
if they correspond to the
same Klein-Gordon wavefunction, we find that
the final
state obtained here is physically distinct from the one obtained
with the original choice of time slicing. Thus, this example -- which is
not clouded by factor ordering ambiguities --
explicitly shows the dependence of the deparametrization method upon
the choice of time slicing.
It is also possible to show slice dependence of the deparametrization method
with fixed initial and final time slices in lowest order in perturbation in
the deformation of slices with a natural factor ordering
\cite{HIG}.

Next, we study the model (\ref{clham}) using the quantization
procedure outlined in
Sec.\ \ref{Prop}. We shall study the dynamics predicted by this
procedure by expressing the theory in $L^2$ form as described
at the end of Sec.\ \ref{Prop}, using the time slicing $t = x_0$,
so that a direct
comparison can be made with the deparametrization procedure. As we
shall see, a ``memory effect" is present: The dynamics defined
by $H_{\it eff}$ does not return to standard, free particle dynamics
after the potential has been turned off,
i.e., the system has the ``memory" that
it had a nonzero potential in the past.  Although this effect will not
be relevant for the cosmological models that
we will study in this paper, it could be important, for
example, in
models in which the universe tunnels through a potential barrier in
an early stage.

In order to calculate $H_{\it eff}$, we must calculate
the operator $C_t$ defined by Eq.\ (\ref{Ct}) above. Since by Eq.\ (\ref{D})
the Klein-Gordon solutions in the region $x_0 > x_0^{(M)}$
which are asymptotically
positive frequency in the past take the form (to first
order in $\lambda$)
\begin{equation}
\Phi  =  e^{-ipt + i{\bf p}\cdot{\bf x}} -i \lambda \int
\frac{d^3{\bf p}'}{(2\pi)^32p'}
\left[ \tilde{V}(p' - p,{\bf p}'-{\bf p})
e^{-ip't+i{\bf p}'\cdot {\bf x}}
-\tilde{V}(-p' - p,{\bf p}'-{\bf p})
e^{ip't + i{\bf p}'\cdot {\bf x}}\right]\,,
\end{equation}
we find
for $x_0 > x_0^{(M)}$
\begin{equation}
C_t e^{i{\bf p}\cdot{\bf x}}  =  p e^{i{\bf p}\cdot{\bf x}}
  -i\lambda \int\frac{d^3{\bf p}'}{(2\pi)^3}
\tilde{V}(-p'- p,{\bf p}'-{\bf p})
e^{i(p + p')t + i{\bf p}'\cdot{\bf x}}\,.
\end{equation}
It follows that to first order in $\lambda$ the operator
$B_t = [C_t + C^{\dagger}_t]^{1/2}$ is given by
\begin{eqnarray}
B_t e^{i{\bf p}\cdot{\bf x}}  & = & \sqrt{2p}e^{i{\bf p}\cdot{\bf x}} \nonumber
 \\
 & & + i\lambda
\int \frac{d^3{\bf p}'}{(2\pi)^3}
\frac{\tilde{V}(p'+p,{\bf p}'-{\bf p})
e^{-i(p + p')t}
 - \tilde{V}(-p'- p,{\bf p}'-{\bf p})
e^{i(p + p')t}}{\sqrt{2p} + \sqrt{2p'}} e^{i{\bf p}'\cdot{\bf x}}\,. \nonumber
\\
  & &
\end{eqnarray}

Since $N = 1$ and $N^a = 0$,
for $x_0 > x_0^{(M)}$
we obtain from Eq.\ (\ref{Heff}) the result
\begin{eqnarray}
H_{\it eff} e^{i{\bf p}\cdot{\bf x}} & = & p e^{i{\bf p}\cdot{\bf x}}
\nonumber \\
 & & + i\lambda\int\frac{d^3{\bf p}'}{(2\pi)^3}
\left[ \frac{\sqrt{p}}{\sqrt{p}+\sqrt{p'}}\tilde{V}(p'+p,{\bf p}'-{\bf p})
e^{-i(p'+p)t + i{\bf p}'\cdot{\bf x}} \right. \nonumber \\
 & & \ \ \ \ \ \ \ \ \ \ \ \ \ \ \ \
\left. -\frac{\sqrt{p'}}{\sqrt{p}+\sqrt{p'}}\tilde{V}(-p'-p,{\bf p}'-{\bf p})
e^{i(p'+p)t + i{\bf p}'\cdot{\bf x}} \right]\,.
\end{eqnarray}
It is readily apparent that $H_{\it eff} \neq H_{\it ADM}$.
Furthermore, although
$H_{\it eff}$ agrees with the standard, free particle Hamiltonian
$H_0 = \sqrt{- \nabla^2}$ prior to time when the potential is ``turned on",
we have $H_{\it eff} \neq H_0$ after the potential is ``turned off". This means
that, in principle, by observing the dynamics of ``free"
particles (after the potential has been turned off), one could deduce that
a potential had previously been present. This
``memory" phenomenon can occur because the positive frequency
condition in the asymptotic past enters the definition of the
operator $C_t$ -- and, thereby, $H_{\it eff}$ -- thus making these operators
nonlocal in time.

Some insight into the nature of this memory effect can be obtained
by considering the simpler case where the potential is
purely a function of time, $t$ (and, thus, is not of compact spacetime
support). Then the positive frequency solution which in the past
(before the potential is turned on) behaves as
$e^{-ip t + i{\bf p}\cdot{\bf x}}$ evolves in the future
(after the potential is turned off) to a solution of the form
\begin{equation}
e^{-ipt + i{\bf p}\cdot{\bf x}} \longrightarrow
\alpha({\bf p})e^{-ipt + i{\bf p}\cdot{\bf x}} +
\beta ({\bf p})e^{+ipt +  i{\bf p}\cdot{\bf x}}\,,
\end{equation}
where $|\alpha({\bf p})|^2 - |\beta({\bf p})|^2 = 1$.  Then, it can
be readily seen that
for $x_0 > x_0^{(M)}$
the effective Hamiltonian is
\begin{equation}
H_{\it eff}e^{i{\bf p}\cdot{\bf x}}
= \frac{p}{|\alpha({\bf p})e^{-ipt} + \beta({\bf p})e^{ipt}|^2}
e^{i{\bf p}\cdot{\bf x}}\,,
\end{equation}
i.e., after the potential has been turned off,
the $L^2$ wavefunctions evolve as
\begin{equation}
\Psi_{L^2} = \frac{\alpha({\bf p})e^{-ipt} + \beta({\bf p})e^{ip t}}
{|\alpha({\bf p})e^{-ipt} + \beta({\bf p})e^{ipt}|}e^{i{\bf p}\cdot{\bf x}}
\,.
\end{equation}
In this case, the phase acquired in
the period $T = 2\pi/p$ is still $-2\pi$ since
$|\alpha({\bf p})| > |\beta({\bf p})|$.  This implies that
for a wavepacket sharply peaked around ${\bf p}$, the motion of
the expectation value of the position operator, ${\bf x}$, will
simply oscillate about
the standard free-particle value.  However,
more complicated dynamical behavior would occur in the model
considered above where $V$ has compact spacetime support.
\end{section}
\setcounter{equation}{0}

\begin{section}{Homogeneous cosmologies}
\label{Hc}
In this section, we will provide more details concerning the application
of the proposal outlined in Sec.\ \ref{Prop} to
general homogeneous cosmological models with a scalar field \cite{COM0}.
The possible homogeneous cosmologies are comprised by
the Bianchi models and the
Kantowski-Sachs model (see, e.g., \cite{MCLM}).  We first consider the
Bianchi models, following the discussion and
notation of Ref.\ \cite{WTEX}. In the Bianchi models, the spacetime
manifold, $M$, is taken to be
$\reell \times G$,
where $G$
is a three dimensional Lie group.
(We parametrize $\reell$ by the variable, $t$.)
For any $k \in G$, we define a diffeomorphism
$\psi_{k}$ -- called left translation by $k$ --
by $\psi_{k} (g) \equiv kg$ for all
$g \in G$. A tensor field ${T^{a...}}_{b...}$ is said to be
left invariant if $\psi_{k}^{*}{T^{a...}}_{b...} = {T^{a...}}_{b...}$,
where $\psi_{k}^{*}$
is the map on tensor fields induced by
$\psi_{k}$.
The spacetime metric on $M$ is taken to be of the form
\begin{equation}
g_{ab} = -\nabla_{a}t \nabla_{b} t + h_{ab}(t) \,,
\end{equation}
where the spatial metric, $h_{ab}$, on $G$ is left invariant, and thus can
be expanded as
\begin{equation}
h_{ab} = \sum_{i,j=1}^{3} h_{ij}(t)
(\sigma^{i})_{a}(\sigma^{j})_{b}\,,
\end{equation}
where $(\sigma^{i})_a$ are left invariant one-forms that are
independent of $t$.

The structure constant tensor field $C^{c}_{\ ab}$ on $G$ is defined by
$[v,w]^c = C^{c}_{\ ab}v^a w^b$
for any two left invariant vector fields $v^a$ and $w^a$.
It is known that $C^{c}_{\ ab}$ can be expressed as \cite{E}
\begin{equation}
C^{c}_{\ ab} = M^{cd}\epsilon_{dab} + \delta^{c}_{[a}A_{b]} \,,
\label{eps}
\end{equation}
where $M^{ab}$ is a left invariant symmetric tensor
on $G$, $A_a$ is a left invariant
one-form and $\epsilon_{abc}$ is a left invariant
three-form on $\Sigma$.
The Jacobi identity implies that $M^{ab}A_b = 0$.
The Lie algebra is said to be of class A if $A_a = 0$, and of class B
otherwise.
No consistent Hamiltonian formulation is known for the Bianchi models
of class B \cite{MAC}.
In particular, the
Hamiltonian constraint does not generate the dynamics of
the system.  We specialize to the models of class A for this reason.

The Lie algebras of class A
are uniquely determined
up to isomorphisms by the rank and the signature of the tensor
$M^{ab}$, where the overall sign of $M^{ab}$
is irrelevant.  Thus, one has
the following six inequivalent cases:
$$
(0,0,0), (+,0,0), (+,-,0), (+,+,0), (+,+,-), (+,+,+).
$$
The corresponding Lie algebras are called the Bianchi types I, II, VI$_{0}$,
VII$_{0}$, VIII and IX, respectively. The unique connected, simply
connected Lie group $G$ for Bianchi type IX is $SU(2)$, which has topology
$S^3$. The corresponding simply connected Lie groups for the other
Bianchi types are noncompact, and only the (trivial) Bianchi I algebra
is also the Lie algebra of a compact Lie group (namely
$U(1) \times U(1) \times U(1)$) \cite{HOSO}.

In the compact case, we choose the configuration variable, $h_{ab}$,
for the gravitational degrees of freedom to be
the spatial metric at the identity element, $e$. The conjugate
momentum variable $\pi^{ab}$, is then the usual momentum
density at $e$ multiplied by the ``volume of space", ${\cal V}$,
as determined by the $3$-form $\epsilon_{abc}$
introduced in eq.(\ref{eps}) (so that $\pi^{ab}$ may be viewed as
the integral of the momentum density over space).
Similarly, we choose the configuration variable for the scalar field
to be the value of $\phi$
at the identity element and its conjugate momentum, $\pi_{\phi}$,
is taken to be the usual
momentum density multiplied by ${\cal V}$.
The momentum constraint reads
\begin{equation}
P_d \equiv \pi^{a}_{\ c}M^{cb}\epsilon_{dab} = 0\,. \label{mom}
\end{equation}
The Hamiltonian constraint is \cite{W2}
\begin{equation}
H \equiv {\cal V}^{-1}h^{-1/2}\left( \pi^{ab}\pi_{ab} -\frac{1}{2}
\pi^2\right) + {\cal V}h^{-1/2}\left( M_{ab}M^{ab}-\frac{1}{2}M^2\right)
+ \frac{1}{24{\cal V}}h^{-1/2}\pi_{\phi}^2 +
\sqrt{24}{\cal V}^2 h^{1/2}V_{s}(\phi) = 0\,,
\label{ham}
\end{equation}
The factors
for the scalar contribution have been chosen for later
convenience. Equations (\ref{mom}) and (\ref{ham}) also hold in the
noncompact case (with ${\cal V}$ an arbitrary constant)
after suitable (infinite)
rescalings have been made on the dynamical variables
and $H$.

In Appendix A, we shall consider the quantum theory obtained by treating
the constraints (\ref{mom}) and (\ref{ham}) on an equal footing.
In particular, we shall derive the explicit form of the Wheeler\--DeWitt
equation for the Bianchi IX model which results from imposing
(\ref{mom}) as an operator constraint on the wavefunction.
However, in the body of this paper we shall consider the models obtained
by first reducing the problem at the classical level -- thereby
eliminating the momentum constraints classically -- and then imposing
(\ref{ham}) as an operator constraint.
We achieve this reduction by noting that the classical momentum
constraint (\ref{mom})
is equivalent to the statement that at each time, $t$,
the quantities
${\pi^a}_b$ and ${M^a}_b$ commute
when viewed as linear maps acting on the tangent space,
$V_e$, at the identity element $e\ \in\ G$.
(Here, all indices are raised and lowered with $h_{ab}$ and $h^{ab}$.)
This means that we can simultaneously diagonalize these linear maps in an
orthonormal basis of $h_{ab}$.
The classical evolution equations then imply that $\pi^{ab} , ~ M^{ab}$,
and $h_{ab}$ will remain diagonal in this basis for all time.
Thus, classically, there is no loss of generality in restricting to the
``diagonal case'',
in which case the momentum constraints become trivial.

It is convenient to choose the diagonal basis $(\sigma^i)_a$ so that $M_i
\equiv M^{ab} (\sigma^i)_a (\sigma^i)_b$ takes the form
\begin{eqnarray}
(M_1,M_2,M_3) & = & (0,0,0), (+1,0,0), (+1,-1,0), \nonumber \\
 & & (+1,+1,0), (+1,+1,-1),
(+1,+1,+1), \nonumber
\end{eqnarray}
respectively, in the six cases corresponding to the different possible
choices of signature.
We then write the basis components of $h_{ab}$ and $\pi^{ab}$ as
$h_{ij} \equiv {\rm diag}(h_1,h_2,h_3)$,
$\pi^{ij} \equiv {\rm diag}(\pi^1,\pi^2,\pi^3)$  and
$\tilde{h}_i \equiv h^{-1/3}h_i$.
We define $\alpha$, $\beta_+$, and $\beta_-$ by
\begin{eqnarray}
h_1 & \equiv & e^{2\alpha + 2(\beta_{+} + \sqrt{3}\beta_{-})}\,, \\
h_2 & \equiv & e^{2\alpha + 2(\beta_{+} - \sqrt{3}\beta_{-})}\,, \\
h_3 & \equiv & e^{2\alpha -4\beta_{+}}\,.
\end{eqnarray}
(Note that this definition of $\alpha$ is consistent with Eq.\ (\ref{alpha}),
i.e., the maps on mini\-superspace defined by Eq.\ (\ref{alpha}) correspond
to a translation of the parameter $\alpha$ defined by the above
equations.) In terms of these variables,
the Hamiltonian constraint becomes
\begin{equation}
H = \left(\frac{{\cal V}}{\sqrt{24}}\right)^{1/2}
e^{-3\alpha} (\tilde{K} + \tilde{P}) = 0\,, \label{const}
\end{equation}
where
\begin{eqnarray}
\tilde{K} & = & -\pi_{\alpha}^2 + \pi_{+}^2 + \pi_{-}^2 +
\pi_{\phi}^2 \\
\tilde{P} & = &
e^{4\alpha}\left[ \frac{1}{2}\left( M_1^2\tilde{h}_{1}^2
+ M_2^2\tilde{h}_2^2 + M_3^2\tilde{h}_3^2\right) \right. \nonumber \\
 & &\ \ \ \ \left. -M_2 M_3 \tilde{h}_1^{-1} - M_3 M_1 \tilde{h}_2^{-1}
-M_1 M_2 \tilde{h}_3^{-1}\right] \nonumber \\
 & & + e^{6\alpha}V_{s}(\phi)\,,
\label{60}
\end{eqnarray}
where $\pi_{\alpha}$, $\pi_{\pm}$ and $\pi_{\phi}$ are the
conjugate momenta of $\alpha$, $\beta_{\pm}$ and
$\phi$, respectively.  We also have eliminated the
factor of $24 {\cal V}^2$ in $\tilde{P}$ by shifting $\alpha$ by
$(1/2) \ln (\sqrt{24}{\cal V})$.
The Wheeler-DeWitt equation corresponding to (\ref{const}) then is
simply a Klein-Gordon
equation in flat spacetime with potential $\tilde{P}$, i.e.,
\begin{equation}
\left[ \partial_{\alpha}^2 - \partial_{\beta_{+}}^2 - \partial_{\beta_{-}}^2
-\partial_{\phi}^2 + \tilde{P} \right] \Phi = 0\,. \label{feq}
\end{equation}

In the Kantowski-Sachs model \cite{KS}, space has the topology
of $S^1 \times S^2$ acted upon by the isometry group $U(1) \times SO(3)$.
After performing a similar, classical reduction of the
momentum constraints, the spacetime metric of this model can be
written in the form
\begin{equation}
ds^2 = -dt^2 + e^{2\left[ \alpha(t) + 2\beta(t)\right]}d\psi^2
+ e^{2\left[ \alpha(t) -\beta(t)\right]}
(d\theta^2 + \sin^2\theta d\phi^2 )\,.
\end{equation}
With scalar field matter present, the rescaled super-Hamiltonian
takes the form (after some redefinition of variables)
\begin{equation}
H = -\pi_{\alpha}^2 + \pi_{\beta}^2 + \pi_{\phi}^2
-e^{4\alpha + 2\beta} + e^{6\alpha}V_s (\phi) = 0\,,
\end{equation}
so the Wheeler-DeWitt equation is simply
\begin{equation}
\left[ \partial_{\alpha}^2 - \partial_{\beta}^2 - \partial_{\phi}^2
 + \tilde{P} \right] \Phi = 0\,,
\label{KSeq}
\end{equation}
\begin{equation}
\tilde{P} =
-e^{4\alpha + 2\beta} + e^{6\alpha}V_s (\phi)\,.
\end{equation}

It is worth elucidating the origin of the differences occurring between
the quantum theory obtained by the above classical reduction of the
momentum constraints
as compared with the method described in the Appendix A.
For definiteness, we focus attention upon the cases of the vacuum
Bianchi I and Bianchi IX models.
In the Bianchi I case, we have $M^{ab} = 0$, so the momentum constraint
is absent entirely.
The above ``classical reduction" to the diagonal case is closely analogous
to formulating a quantum theory of a free, nonrelativistic particle
moving in $\reell^3$ by first noting that, classically, the particle moves
in a straight line, so that, classically, we may choose our
coordinates in $\reell^3$ so that
the particle moves only in the $x$\--direction.
We may then formulate a quantum theory equivalent to that of a
free particle moving in $\reell^1$. However,
this reduced theory --  analogous to the
treatment of the Bianchi I model given above --
is not equivalent to standard free particle theory in $\reell^3$,
which is the analog of the
treatment of the Bianchi I model given in Appendix A. On the other hand,
for the Bianchi IX case, the momentum constraints are nontrivial, and the
number of degrees of freedom are the same for the above classically
reduced theory as for the theory obtained via a quantum reduction of the
momentum constraints as given in Appendix A.
A good analog of the Bianchi IX model is a free, nonrelativistic particle
in $\reell^3$ with the imposition of the additional constraint that its
vector angular momentum vanishes. The analog of above
classical reduction
procedure would be to note that, classically, the particle moves on
a straight line through the origin, and then to formulate a quantum
theory equivalent to that of a free particle moving in $\reell^1$. The
analog of the procedure given in Appendix A would be to solve the
constraint by restricting attention to wavefunctions in $\reell^3$ which
are spherically symmetric. This yields a theory which is not unitarily
equivalent to standard free particle theory in $\reell^1$ \cite{RTS}.

Clearly, the quantum treatment of the momentum constraints
given in Appendix A is
``more correct" than the classical reduction of them given above.
Nevertheless, in the body of this paper
we choose to work with the models obtained by the above classical
reduction, since the Wheeler-DeWitt equation is considerably simpler
in this case --
being a wave equation in a flat rather than curved spacetime. In
addition, most of the previous literature has considered the
Bianchi models
obtained by the classical reduction, so our consideration of them here
should facilitate comparison with previous approaches. Since the
models will be used only to elucidate the qualitative features of the
quantum theory proposed in Sec.\ \ref{Prop} -- rather than
to attempt to make realistic predictions about
the actual universe -- we see little disadvantage to working with the
simpler, classically reduced models.

We turn our attention, now, to an investigation of whether the
following two conditions -- which are necessary and sufficient
for the implementation of the proposal of Sec.\ \ref{Prop} -- are
likely to be satisfied in the above Bianchi and Kantowski-Sachs models:
(1) For the construction of the Hilbert space,
${\cal H}$, of states, it is necessary for
the potential term in the Wheeler-DeWitt equation to vanish sufficiently
rapidly as $\alpha \to - \infty$ that the notion of ``asymptotically
positive frequency solutions" is well defined. (2) For the construction
of observables on a ``time slice" $\Sigma_t$, it is necessary that the
space, ${\cal D}$,
of $C^1$ solutions in ${\cal H}$ whose restriction to $\Sigma_t$ lies
in $L^2(\Sigma_t)$ be dense in both ${\cal H}$ and $L^2(\Sigma_t)$; it
also is necessary that any sequence in ${\cal D}$ which converges in both
${\cal H}$ and $L^2(\Sigma_t)$ have a nonzero limit in ${\cal H}$ if and
only if it has a nonzero limit in $L^2(\Sigma_t)$ \cite{W}.

With regard to condition (1), it should be noted that in all of our models
(as well as in full quantum gravity), the potential terms vanish
exponentially rapidly as $\alpha \to - \infty$. Nevertheless, since the
potential terms also may blow up exponentially rapidly at spatial infinity,
it does not automatically
follow that these terms can be neglected at ``early times", in the manner
needed for the construction of ${\cal H}$.

The following criterion is very useful
to consider for the analysis of condition (1):
Consider the vector space, $S$, of real
solutions to the Wheeler-DeWitt equation with initial data of compact
support on Cauchy surfaces. Define an ``energy inner product"
on $S$ associated with each Cauchy surface $\alpha= {\rm const.}$ by
\begin{equation}
E_\alpha(\Phi_1,\Phi_2) \equiv
\int_{\alpha= {\rm const.}} d\Sigma^A
\left[ \partial_A \Phi_{1}\partial_B \Phi_2
+ \frac{1}{2} \tilde{G}_{AB} (\partial_C \Phi_1\partial^C\Phi_2 +\tilde{P}
\Phi_1\Phi_2) \right](\partial/\partial \alpha)^B \,,
\end{equation}
where $\tilde{G}_{AB}$ is the flat metric appearing in
(\ref{feq}) and (\ref{KSeq}),
so that the ``energy norm"
at ``time" $\alpha$ of a solution, $\Phi$, is just the integral over the
surface at that value of $\alpha$ of
the stress-energy tensor of $\Phi$ contracted once with the unit normal
and once with $\partial / \partial \alpha$.
Then the following condition is sufficient
to enable the desired Hilbert space, ${\cal H}$, to be constructed:

\noindent
{\em Condition (1$'$)}: For all $\Phi_1,\Phi_2 \in S$ the limit
\begin{equation}
E(\Phi_1, \Phi_2) \equiv
\lim_{\alpha\to -\infty}E_\alpha(\Phi_1,\Phi_2)
\end{equation}
exists and satisfies
\begin{equation}
E(\Phi_1, \Phi_1) E(\Phi_2, \Phi_2) \ge
K \vert \Omega(\Phi_1,\Phi_2)\vert^2
\end{equation}
for some constant $K > 0$, where $\Omega$ was defined by Eq.\ (\ref{2inn}).

That condition (1$'$) implies that the desired ${\cal H}$
can be constructed
follows immediately from the work of Chmielowski \cite{CH}.
We define the inner product, $\mu$, on $S$ satisfying
\begin{equation}
\mu(\Phi_1,\Phi_1) = \ ^{\rm LUB}_{\Phi_2 \neq 0}
\frac{\vert \Omega(\Phi_1,\Phi_2)\vert^2}{4\mu(\Phi_2,\Phi_2)}
\end{equation}
to be the inner product associated to $E$ by the construction of
Proposition 1 of \cite{CH}. A Hilbert space, ${\cal H}$, of solutions
then can be constructed from $\mu$ as discussed in detail in, e.g.,
\cite {KW} and \cite{WBOOK}.
This Hilbert space has the interpretation of being comprised of the
``asymptotically positive frequency solutions" for the same reason
that the standard one-particle Hilbert space for stationary spacetimes
\cite{AMK} has the interpretation of being comprised of positive
frequency solutions.

In the vacuum case, we believe that condition (1$'$) holds
for the Bianchi models I, II, VI$_{0}$, and VII$_{0}$ as well as for
the Kantowski-Sachs model. However, it appears that condition (1$'$) will fail
for the vacuum Bianchi VIII and IX models. Nevertheless,
we believe that condition (1$'$) will hold
in all of the homogeneous cosmological models (including VIII and IX)
in which a scalar field, $\phi$, is present, provided only that the
potential, $V(\phi)$,
does not grow exponentially rapidly in $\phi$.

Our arguments for these beliefs are based
mainly upon the classical ``particle"
dynamics associated with the super-Hamiltonian
(\ref{const}), since the behavior
of suitable ``wavepackets" satisfying (\ref{feq}) should be similar to this
classical particle dynamics in the limit $\alpha \to - \infty$. Consider,
first,
the vacuum case.
The spatial region in the $(\beta_+{\rm -}\beta_-)$-plane where the terms
\begin{eqnarray}
 & & e^{4\alpha}(-M_2 M_3 \tilde{h}_1^{-1} - M_3 M_1 \tilde{h}_2^{-1} -
M_1 M_2 \tilde{h}_3^{-1} ) \nonumber \\
 & & = -M_2 M_3 e^{4\left[ \alpha - (1/2)(\beta_{+}
+ \sqrt{3}\beta_{-})\right]}
-M_3 M_1 e^{4\left[ \alpha
- (1/2)(\beta_{+} - \sqrt{3}\beta_{-})\right]}
-M_1 M_2 e^{4(\alpha + \beta_{+})} \nonumber
\end{eqnarray}
in $\tilde{P}$ of (\ref{60}) are large ``moves away" from
the origin at the speed of light as $\alpha \to -\infty$.
Hence, for large enough $-\alpha$, these terms should be
negligible in both the particle and wave dynamics, and the relevant
contribution from $\tilde{P}$ should be
\begin{eqnarray}
\tilde{P} & \approx & \frac{1}{2} ( M_1^2 \tilde{h}_1^2
+ M_2^2 \tilde{h}_2^2 + M_3^2\tilde{h}_3^2 ) \nonumber \\
 & = & \frac{M_1^2}{2} e^{4\alpha + 4\beta_{+} + 4\sqrt{3}\beta_{-}}
  + \frac{M_2^2}{2} e^{4\alpha + 4\beta_{+} - 4\sqrt{3}\beta_{-}}
  + \frac{M_3^2}{2} e^{4\alpha -  8\beta_{+}}\,.
\end{eqnarray}
These terms are well approximated as
$\alpha \to -\infty$ by potential walls located, respectively, at
$\beta_{+} + \sqrt{3}\beta_{-} = -\alpha$,
$\beta_{+} - \sqrt{3}\beta_{-} = -\alpha$ and $-2\beta_{+} = -\alpha$
\cite{COM5}.
These walls surround a region in the shape of a triangle and
recede from the origin at half the speed of light.
In Bianchi types I, II, VI$_{0}$, and VII$_{0}$,
the rank of $M^{\alpha\beta}$
is less than three, i.e., one or more of the $M_i$'s are zero, so at least
one of these walls will be missing. In that case,
if we consider evolution backwards in $\alpha$, a generic particle
trajectory should
bounce from the remaining walls at most a finite
number of times, after which time it can be treated as a free particle.
Similarly, wavepackets should, after a finite time, escape to a
region where they satisfy the free Klein-Gordon equation
to an excellent approximation.
Condition (1$'$) should then be satisfied
in the vacuum Bianchi I, II, VI$_{0}$, and VII$_{0}$ models.
(Indeed, these arguments can be made rigorous for the vacuum Bianchi I and II
models, since they can be solved explicitly.)
Similar arguments apply (rigorously)
to the vacuum Kantowski-Sachs model, where
the region where the potential, $-e^{4\alpha + 2\beta}$, is large recedes
from the origin at twice the speed of light.

In Bianchi type VIII and IX models we have
$M_1^2 = M_2^2 = M_3^2 = 1$, so all the ``potential walls" are present.
Since these walls recede only at
half the speed of light --  whereas in the classical dynamics,
the ``particle'' in the mini\-superspace always moves at the speed of
light between ``bounces" -- an infinite number of ``bounces" generically
occurs as $\alpha \to -\infty$ \cite{MIX}. Since energy is lost on
each of the ``bounces" the energy should asymptotically approach zero.
Similarly, in the wave dynamics, it appears that $E_\alpha \to 0$ as
$\alpha \to -\infty$, and condition (1$'$) should fail to hold.

However, the situation improves considerably if scalar field
matter is included, provided that $V(\phi)$
does not grow exponentially rapidly in $\vert \phi \vert$. In that
case, the scalar potential term,
$e^{6\alpha}V(\phi)$,
should become negligible for large $-\alpha$, and
the momentum $\pi_{\phi}$ should become approximately conserved.
The classical dynamics as $\alpha \to -\infty$ then corresponds to that
of a massive particle,
with mass $\pi_{\phi}^2$. Although the potential walls still recede at
only half the speed of light, on account of the mass term
the classical particle now
will slow down each time it bounces off a wall.  Eventually, its velocity
will drop below
half the speed of light, at which point the particle will no longer
see the potential walls. Thus, if $\pi_{\phi} \ne 0$, only a finite number of
bounces will occur, and the classical dynamics will not be chaotic as
$\alpha \to -\infty$, i.e., the presence of a scalar field is sufficient to
eliminate the
classical chaotic behavior of the Bianchi VIII and IX models \cite{HI}.
Furthermore, the energy of classical solutions
will be approximately conserved as $\alpha \to -\infty$. Similar behavior
should occur for wavepackets, so it appears that
condition (1$'$) should hold.
Thus, it seems likely
that all of the
homogeneous cosmological models with a scalar field will satisfy the
first condition needed for the implementation of the
proposal of Sec.\ \ref{Prop}. Note, however,
that it is known that the classical chaotic behavior is not
eliminated if one includes
a homogeneous electromagnetic field rather than a scalar field in
Bianchi type IX model \cite{WALL}, so it appears that
a scalar field is a necessary ingredient for the construction of the
Hilbert space in the Bianchi VIII and IX models.

As for condition (2) -- which
requires that there be a dense subset of
$L^2$-normalizable functions in the Hilbert
space -- we see no reason to doubt its validity
in the Bianchi models for the surfaces of constant $\alpha$.
However, in the Kantowski-Sachs model the
solutions to Eq.\ (\ref{KSeq}) in the vacuum case which can be obtained by
separation of the variables $\tau = (2\alpha +\beta)/\sqrt{3}$,
$\xi = (2\beta +\alpha)/\sqrt{3}$
turn out to grow faster than
exponentially for large
$\beta$. We expect condition (2) to hold on the surfaces of
constant $\tau$, but the rapid growth of the
solutions in $\beta$ suggests that
condition (2) will fail for the surfaces of constant $\alpha$,
although we have not been able to prove this. Similarly,
in the Bianchi IX model (as well as
models obtained by suppressing some of its degrees of freedom, such as those
studied in the next two sections), it appears that condition (2) will fail
for some choices of Cauchy surface other than the surfaces of constant
$\alpha$.
\end{section}
\setcounter{equation}{0}

\begin{section}{Robertson-Walker universe with a scalar field}
\label{Rw}
In this section we apply the quantization prescription of Sec.\ \ref{Prop} to
the closed Robertson-Walker cosmology with a
homogeneous, free, massless scalar field.
We shall choose
the hypersurfaces of constant $\alpha$ as our time slices
in mini\-superspace.

The classical (rescaled)
super-Hamiltonian can be
obtained from that of the Bianchi IX model, Eq.\ (\ref{const}), by
setting $\pi_{\pm} = \beta_{\pm} = 0$, i.e.,
\begin{equation}
H_{RW} = -\pi_{\alpha}^2 + \pi_{\phi}^2 - e^{4\alpha} = 0 \,.
\label{HRW}
\end{equation}
The trajectories in mini\-superspace corresponding to the classical solutions
of the equations of motion for this super-Hamiltonian are given by
\begin{equation}
\alpha  = \frac{1}{2}\left\{ \ln |p| - \ln\cosh\left[ 2(\phi -\phi_0)\right]
\right\} \,,
\label{classtraj}
\end{equation}
where $\phi_0$ and $p$ $(=\pi_{\phi})$
are arbitrary real constants.
These trajectories describe classical universes which start at a ``big bang"
singularity at $\alpha \to -\infty$ (with $|\phi| \to \infty$), expand to
the maximum size
$\alpha_{M} = (1/2)\ln |p|$ at $\phi = \phi_0$, and then recollapse
in a symmetrical manner.

We have two main motivations for studying this model: (i) As noted above,
all of the classical solutions recollapse. On the other hand, the proposal
of Sec.\ \ref{Prop} treats $\alpha$ as a ``time variable", which can be
prescribed arbitrarily to ``set the conditions" for the other dynamical
variables. Thus, in particular, we could choose a state in ${\cal H}$
which, at $\alpha \to -\infty$, corresponds closely to a classical trajectory
(\ref{classtraj}), and ask about the behavior of $\phi$ for
$\alpha \gg \alpha_M$.
How does $\phi$ behave in this classically
forbidden region of mini\-superspace? (ii) The
classical trajectory (\ref{classtraj}) is everywhere spacelike in
mini\-superspace. On the other hand, the propagation defined by
the Wheeler-DeWitt equation is entirely causal with respect to the light
cones defined by the metric on mini\-superspace. Thus, one might anticipate
some difficulties in approximating a classical trajectory by a state in
${\cal H}$ even during the ``expanding phase" of the classical solution.
Clearly, the states in ${\cal H}$ do not behave classically for
$\alpha > \alpha_M$. How close can one get to $\alpha_M$ before the
universe begins to behave nonclassically?

As we now shall see, for wavepackets in ${\cal H}$ corresponding to
classical solutions which expand for much longer than the Planck time,
the answer to question (i) is that for $\alpha > \alpha_M$, $\phi$ ``freezes"
at the value, $\phi_0$, corresponding to the
classical value of $\phi$ at maximum expansion, $\alpha = \alpha_M$.
The answer to question (ii) is that despite the spacelike character of the
classical trajectory, suitably chosen wavepackets in ${\cal H}$ accurately
describe the expanding phase of the classical solution until
$\alpha \approx \alpha_M$.

The Wheeler-DeWitt equation for this model can be obtained by
setting $\pi_{\alpha} = -i\partial_{\alpha}$
and $\pi_{\phi} = -i\partial_{\phi}$ in Eq.\ (\ref{HRW}), i.e., we have
\begin{equation}
H_{RW}\Phi = \left[\frac{\partial^2\ }{\partial\alpha^2}  -
\frac{\partial^2\ }{\partial\phi^2} - e^{4\alpha}
\right]\Phi(\alpha,\phi) = 0\,.
\label{WDWRW}
\end{equation}
Note that the deparametrization approach
of Sec.\ \ref{Adm}
cannot be employed here because the
potential is negative, but the method of Sec.\ \ref{Prop} can be
used without difficulty.

Since  the momentum $\pi_{\phi}$ is conserved, the solutions to the equation
$H_{RW}\Phi = 0$ can be written as
$\Phi_p(\alpha,\phi) \propto f_{|p|}(\alpha)e^{ip\phi}$,
where the function $f_{|p|}(\alpha)$ satisfies
\begin{equation}
\left(\frac{d^2\ }{d\alpha^2} + p^2 - e^{4\alpha}\right)f_{|p|}(\alpha)
= 0\,. \label{3eig}
\end{equation}
The two linearly independent solutions of this equation are
$I_{\pm i|p|/2}(e^{2\alpha}/2)$,
where
\begin{equation}
I_{\nu}(z) \equiv \sum_{k=0}^{\infty}\frac{1}{k!\Gamma(\nu + k + 1)}
\left( \frac{z}{2}\right)^{\nu+2k}\,.
\end{equation}
The asymptotic positive frequency condition requires that
$f_{|p|}(\alpha) \to e^{-i|p|\alpha}$ for
$\alpha \to -\infty$.  Thus, the solutions relevant for
constructing wavepackets which lie in ${\cal H}$ are
\begin{equation}
f_{|p|}(\alpha) = 2^{-i|p|}\Gamma(1 - i|p|/2) I_{- i|p|/2}(e^{2\alpha}/2)\,.
\end{equation}
If we impose the $\delta$-function normalization
$\langle \Phi_p,\Phi_{p'} \rangle_{KG} =\delta(p-p')$, our ``basis"
for solutions
in ${\cal H}$ is
\begin{equation}
\Phi_{p}(\alpha,\phi) = \frac{f_{|p|}(\alpha)}{\sqrt{2|p|}}
\frac{e^{ip\phi}}{\sqrt{2\pi}}\,.
\end{equation}

The function $I_{\nu}(z)$ behaves for large $|z|$ as \cite{GR}
\begin{equation}
I_{\nu}(z) \sim \frac{e^{z}}{\sqrt{2\pi z}}\,.
\end{equation}
Hence, in the classically forbidden region $e^{\alpha} \gg |p|$,
the function $f_{|p|}(\alpha)$ grows like
$\exp(e^{2\alpha}/2-\alpha)$.
This behavior of $f_{|p|}(\alpha)$ can be
understood as follows. Eq.~(\ref{3eig}) has the form of
a time-independent Schr\"odinger equation,
\begin{equation}
\left[- \frac{d^2\ }{d\alpha^2} + e^{4\alpha}\right]f_{|p|}(\alpha)
= p^2f_{|p|}(\alpha)\,, \label{3sch}
\end{equation}
with the term $e^{4\alpha}$ acting as a potential barrier.
At large negative $\alpha$ the solutions of this
equation are oscillatory, whereas at
large positive $\alpha$ the solutions are growing and decaying.
If one were solving a scattering problem in Schr\"odinger
quantum mechanics, one would demand that the solution decay at
large positive $\alpha$, in which case one would find equal admixtures
of the oscillating solutions, $e^{-i|p|\alpha}$ and $e^{+i|p|\alpha}$, as
$\alpha \to -\infty$. In our case, however, we demand that the solution
behave as $e^{-i|p|\alpha}$ as $\alpha \to -\infty$, in which case the
growing solution is present (and dominates) as $\alpha \to +\infty$.

The dynamical behavior of $\Phi$ in the classically forbidden region
$e^{\alpha} \gg |p|$ is most easily examined in the $L^2$-version of the
theory described in Sec.\ \ref{Prop}. By Eq.\ (\ref{heffqc}),
the effective Hamiltonian which governs the $\alpha$-evolution of the
$L^2$-wavefunction, $\Psi_p(\alpha,\phi)$, corresponding to
$\Phi_p(\alpha,\phi)$ is
\begin{equation}
H_{\it eff}\Psi_{p}(\alpha,\phi)
= \frac{|p|}{|f_{|p|}(\alpha)|^2}
\Psi_{p}(\alpha,\phi)\,.  \label{he}
\end{equation}
Since the function $f_{|p|}(\alpha)$
grows rapidly for large $\alpha$, the effective
Hamiltonian approaches zero rapidly. This implies that, as claimed above,
the dynamical
evolution of $\phi$ is ``frozen" in the classically forbidden region, i.e.,
the probability distribution for $\phi$ does not change with ``time",
$\alpha$, when $e^{4\alpha} \gg p^2$.

In the region where $e^{4\alpha} \ll p^2$,
the wavefunctions $\Phi_{p}(\alpha,\phi)$ satisfy the
free Klein-Gordon equation to an excellent approximation. Hence, there
is no difficulty in constructing wavepackets which closely follow
the classical trajectory in this region. On the other hand, as we have just
seen, highly nonclassical behavior occurs for $e^{4\alpha} \gg p^2$.
It is of interest to examine the behavior of wavepackets in the classically
allowed region, $e^{4\alpha} < p^2$, when
$e^{4\alpha} \sim p^2$ in order
to determine more precisely where the
semiclassical behavior breaks down.
To investigate this, we consider the
WKB approximation \cite{CK} and analyze the conditions under which
it is valid and gives evolution corresponding closely to the classical
dynamics.
Since the equation we are
solving is equivalent to the Schr\"odinger equation
(\ref{3sch}), we see immediately that the WKB solution to order $\hbar$
which is positive frequency in the asymptotic past is
\begin{equation}
f^{\it WKB}_{|p|}(\alpha) =
\sqrt{\frac{|p|}{k(\alpha,p)}}\exp\left[ -\frac{i}{\hbar}
\int^{\alpha} k(\alpha,p)d\alpha
\right]\,,
\label{fWKB}
\end{equation}
where
\begin{equation}
k(\alpha,p)\equiv \sqrt{p^2 -e^{4\alpha}} \,.
\end{equation}
The corresponding
$L^2$-wavefunction is
\begin{equation}
\Psi^{\it WKB}_{p}(\alpha,\phi) =
\frac{1}{\sqrt{2\pi}}\exp\left[ -i\int^{\alpha} k(\alpha,p)d\alpha +ip\phi
\right]\,.
\end{equation}

The criterion for the validity of the WKB approximation is given by \cite{S}
\begin{equation}
\frac{1}{2k(\alpha,p)^2}\left| \frac{\partial\ }{\partial\alpha}
k(\alpha,p)\right| = \frac{e^{4\alpha}}{\left[ p^2 -e^{4\alpha}\right]^{3/2}}
\ll 1\,.
\end{equation}
This inequality
is satisfied if $p^2 -e^{4\alpha} \gg p^{4/3}$, which, in turn, is satisfied
if $p \gg 1$ and $p^{1/2} - e^{\alpha} \gg p^{-1/6}$.
Restoring
the Planck length, $l_P$, we find $p \sim (l_M/l_P)^2$,
where $l_M$ is the maximum
classical radius of the universe.  Hence, the WKB approximation is valid
provided only that $l_M \gg l_P$ and
$l_M - l(\alpha) \gg (l_P/l_M)^{1/3}l_P$, where
$l(\alpha) \equiv e^{\alpha}l_P$.  Thus, as long as the universe expands
classically to a radius much larger than the Planck length, the WKB
approximation is valid essentially up to the classical turning point.

When the WKB approximation holds, the effective Hamiltonian is given
by Eqs.\ (\ref{he}) and (\ref{fWKB}).
On the other hand, the classical deparametrized
Hamiltonian (see Sec.\ \ref{Adm}) with respect to
the time variable $\alpha$ is simply
$H_{\it cl} = k(\alpha,p)$.
Hence, when the WKB
approximation is valid, we have $H_{\it eff} \approx H_{\it cl}$.
Consequently,
there should be no difficulty in constructing wavepackets which
follow closely the classical trajectories \cite{COM4}.

The above conclusions can be verified numerically.
Fig.\ 1 shows
the quantity $d\phi/d\alpha \equiv \partial H_{\it eff}/\partial p$
as a function of $\alpha$ for $p = 4, 16, 64$ and $256$.
It is compared with the ``true'' classical trajectory, $\phi_{\it cl}$,
given by Eq.\ (\ref{classtraj}).
One can clearly see
that the quantity $d\phi/d\alpha$ is approximated by $d\phi_{cl}/d\alpha$
better and better as $p$ increases.
Note also that
the freezing of dynamics
occurs more and more sharply as $|p|$ increases.  Fig.\ 2 shows
the wavepacket which, for $\alpha \to -\infty$, takes the form
\begin{equation}
\Phi_{p,\Delta p}  =  \left(\frac{2\Delta p}{\sqrt{\pi}}\right)^{1/2}
 \exp\left\{ -2(\Delta p)^2\left[\phi + \frac{1}{2}
\ln 2p_0\right]^2 -ip_0(\alpha - \phi)\right\} \nonumber
\end{equation}
with $p_0 = 64$ and $\Delta p = 10$.
The shaded area represents the interval in $\phi$ for each $\alpha$ which
contains 90\% of the squared $L^2$-normalized wavefunction.
This is compared with the classical solution
$\alpha = (1/2)[\ln p_0 -\ln\cosh(2\phi)]$.
This wavepacket indeed follows the
classical trajectory quite closely up to the classical turning point
at $(\phi,\alpha) = (0, (1/2)\ln p_0)$.
Then it is seen to freeze near $\phi =0$ for
$\alpha > (1/2)\ln p_0$.

Finally, we comment briefly on some differences between the approach
taken here and those taken by Hartle and Hawking \cite{HH,H}
and Vilenkin \cite{V1,V2,V3}.
Our approach defines an entire Hilbert space of states, ${\cal H}$, and
no state vector, $\Phi \in {\cal H}$, is, in any sense ``preferred".
On the other hand, both Hartle and Hawking and Vilenkin seek to
single out a particular wavefunction (without attempting to define a
Hilbert space of states) via the imposition of boundary conditions.
Nevertheless, one may inquire as to whether the Hartle-Hawking or
Vilenkin wavefunctions lie in our Hilbert space, ${\cal H}$.
Even in the context of minisuperspace models, it is not clear how
to implement, in a mathematically precise manner, the general principles
which have been proposed to determine
the Hartle-Hawking and Vilenkin boundary conditions.
However, in the context of the simple model considered in this section,
it seems likely that the Hartle-Hawking wavefunction is the solution to
Eq.\ (\ref{WDWRW})
for which $\Phi$ is independent of $\phi$ and
$\Phi(\alpha) \to  {\rm const.} \neq 0$ for
$\alpha \to -\infty$, i.e., $\Phi_{HH}(\alpha) = I_0(e^{2\alpha}/2)$.
Similarly, the Vilenkin wavefunction
should be the solution to Eq.\ (\ref{WDWRW})
for which $\Phi$ is independent of $\phi$ and
$\Phi(\alpha)$ decays for
$\alpha \to +\infty$, i.e., $\Phi_V(\alpha) = K_0(e^{2\alpha}/2)$.
In any case, since the field transformation $\phi \to \phi + {\rm const.}$
is a symmetry of our model, it is clear that both the Hartle-Hawking
and Vilenkin wavefunctions
(if they exist)
should be independent of $\phi$.
On the other hand, any wavefunction in our
Hilbert space approaches a Klein-Gordon normalizable free-particle
wavefunction for $\alpha \to -\infty$ and the average of $\pi_{\phi}^2$
is necessarily nonzero.  Hence, neither the Hartle-Hawking nor Vilenkin
wavefunctions lie in our Hilbert space.
\end{section}
\setcounter{equation}{0}

\begin{section}{The Taub model}
\label{Taub}
In this section we study the Taub model \cite{ATU}, which
can be obtained from the Bianchi IX model by
letting $\beta_{-} = \pi_{-} = 0$. {}From (\ref{const}) we see that the
classical super-Hamiltonian for the Taub model is
\begin{equation}
H_{T} = -\pi_{\alpha}^2 + \pi_{\beta_{+}}^2
 + e^{4\alpha}V(\beta_{+}) = 0\,,  \label{tsuper}
\end{equation}
where
\begin{equation}
V(\beta_{+}) \equiv \frac{1}{3}e^{-8\beta_{+}}
-\frac{4}{3}e^{-2\beta_{+}}\,.
\end{equation}
As in the Robertson-Walker model of the previous section, all of the
classical solutions start at an initial singularity at $\alpha \to -\infty$,
expand to a maximum size, and then recollapse.

The super-Hamiltonian (\ref{tsuper}) is simplified by
defining
\begin{equation}
\tau = (2\alpha -\beta_{+})/\sqrt{3} \,,
\label{tau}
\end{equation}
\begin{equation}
\xi = (2\beta_{+} -\alpha)/\sqrt{3}
\label{xi}
\end{equation}
so that the new coordinates,
$(\tau, \xi)$, on mini\-superspace differ from $(\alpha, \beta_{+})$
by a Lorentz boost.
In terms of the new coordinates, the Wheeler-DeWitt equation takes
the form
\begin{equation}
H_{T} \Phi = \left[
\frac{\partial^2\ }{\partial\tau^2} - \frac{4}{3}e^{2\sqrt{3}\tau}
- \frac{\partial^2\ }{\partial\xi^2} + \frac{1}{3}e^{-4\sqrt{3}\xi}\right]
\Phi = 0\,.
\end{equation}
This equation can be solved exactly \cite{MR}
by separation of variables. The equations for $\xi$ and $\tau$ both take
the form
\begin{equation}
\left[ \frac{d^2\ }{dx^2} - Ae^{Bx} \right] F(x) = -\omega^2 F(x)\,.
\end{equation}
The general solution to this equation is
\begin{equation}
F(x) = Z_{2i\omega/B}\left( \frac{2\sqrt{A}}{B}e^{\frac{B}{2}x}\right)\,,
\end{equation}
where $Z_{\nu}(z)$ is any modified Bessel function. The relevant solutions
for ${\cal H}$ are those of the form
\begin{equation}
\Phi_{\omega}(\tau,\xi) = \frac{1}{\sqrt{2\omega}}
f_{\omega}(\tau)S_{\omega}(\xi)\,.
\end{equation}
where $f_{\omega}(\tau) \to e^{-i\omega \tau}\ (\omega > 0)$
for $\tau \to -\infty$ and where
$S_{\omega}(\xi)$ has nonsingular $\xi$-dependence. These conditions
determine $f_{\omega}$ and $S_{\omega}$ to be
\begin{equation}
f_{\omega}(\tau) = 3^{-i\omega/\sqrt{3}}\Gamma(1-i\omega/\sqrt{3})
I_{-i\omega/\sqrt{3}}\left(\frac{2}{3}e^{\sqrt{3}\tau}\right)
\end{equation}
and
\begin{equation}
S_{\omega}(\xi) \equiv \sqrt{\frac{\omega\sinh\left[\pi\omega/(2\sqrt{3})
\right]}
{\sqrt{3}\pi^2}}
K_{-i\omega/(2\sqrt{3})}\left(\frac{1}{6}e^{-2\sqrt{3}\xi}\right)\,.
\end{equation}
Here we have normalized $S_{\omega}$ by requiring \cite{HMS}
\begin{equation}
\int_{-\infty}^{+\infty}d\xi S_{\omega}(\xi)S_{\omega'}(\xi)
= \delta(\omega-\omega')
\end{equation}
and we have chosen the phase of $S_{\omega}(\xi)$ to make it
real. With this normalization,
we have $\langle\Phi_{\omega},\Phi_{\omega}'\rangle_{KG} =
\delta(\omega-\omega')$.

In this model, it is easiest to investigate dynamical evolution by choosing
the ``time slices" in mini\-superspace to be the hypersurfaces of constant
$\tau$ rather than $\alpha$. Again, we have $N = 1$ and we choose
$N^a = 0$. The $L^2$-wavefunctions corresponding to
$\Phi_{\omega}(\tau,\xi)$ are given by
\begin{equation}
\Psi_{\omega}(\tau,\xi) \equiv B_\tau \Phi_{\omega}(\tau,\xi) =
\frac{f_{\omega}(\tau)} {|f_{\omega}(\tau)|}S_{\omega}(\xi)\,.
\end{equation}
and, by Eq.\ (\ref{heffqc}), the effective Hamiltonian is
\begin{equation}
H_{\it eff}\Psi_{\omega}(\tau,\xi) = \frac{\omega}{|f_{\omega}(\tau)|^2}
\Psi_{\omega}(\tau,\xi)\,.
\end{equation}
Since $f_{\omega}(\tau)$ blows up for $\tau \to +\infty$,
we find that --  as in the case of the scalar field dynamics in the
Robertson-Walker model -- $H_{\it eff}$ rapidly goes to zero,
and the dynamics of $\xi$ with respect to $\tau$-time
becomes ``frozen" in the classically forbidden region,
$\frac{4}{3}e^{2\sqrt{3}\tau} > \omega^2$.

The WKB approximation can be considered for the $\tau$-dependence of the
wavefunction with fixed $\omega$.
As in the Robertson-Walker model of the previous section, the criterion for
the validity of the WKB approximation is satisfied in the classically
allowed region,
$\frac{4}{3}e^{2\sqrt{3}\tau} < \omega^2$,
until very close to the classical turning point, provided only
that the universe expands to a radius
much larger than the Planck length.  Thus, the
(un-normalizable) basis functions for $L^2$ states
can be approximated by
\begin{equation}
\Psi_{\omega}(\tau,\xi) \approx
\exp\left( -i\int^{\tau}d\tau \kappa(\tau,\omega)\right)S_{\omega}(\xi)\,,
\label{97}
\end{equation}
where
\begin{equation}
\kappa(\tau,\omega) \equiv
\left( \omega^2 - \frac{4}{3}e^{2\sqrt{3}\tau}\right)^{1/2}\,.
\end{equation}
Similar arguments apply
to the function $S_{\omega}(\xi)$.
In this case,
the WKB approximation breaks down near the ``spatial'' classical turning
point, i.e., for $e^{-4\sqrt{3}\xi} \approx 3\omega^2$.  However, as long
as $\omega^2 \gg 1$ , the region where the WKB approximation fails is small.
Thus, except
in a small region near
the (spatial) classical turning point, the function $S_{\omega}(\xi)$ can be
approximated by a linear combination of the functions
\begin{equation}
S^{WKB}_{\omega\pm}(\xi) = \sqrt{\frac{\omega}{2\pi k(\xi,\omega)}}
\exp\left[ \pm i\int^{\xi} d\xi k(\xi,\omega)\right] \,,
\end{equation}
where
\begin{equation}
k(\xi,\omega) \equiv \left( \omega^2 - \frac{1}{3}e^{-4\sqrt{3}\xi}
\right)^{1/3}.
\end{equation}
Combining this result and that for the $\tau$-dependent part [Eq.\ (\ref{97})],
we conclude that
a wavepacket can be made to follow the classical trajectory
(except very near the point where the wavepacket bounces off the ``wall"
in $\xi$) up to the
classical turning point from expansion to contraction.

Again, this conclusion can be verified numerically.
Fig.\ 3 shows a wavepacket with the central
value of $\bar{\omega} \equiv \omega/\sqrt{3}= 32$
and $\Delta\bar{\omega} = 5$
which is arranged in such
a way that it bounces off the wall before reaching the maximum expansion.
It clearly shows that
the wavepacket can follow the classical trajectory up to the classical turning
point in $\tau$ and then freeze there.

The same analysis can be repeated for the Taub model with a homogeneous
scalar field.  The super-Hamiltonian of this model is given by
\begin{equation}
H_{TS} = -\pi_{\alpha}^2 + \pi_{\beta_{+}}^2 + \pi_{\phi}^2
 + e^{4\alpha}V(\beta_{+}) = 0\,.
\end{equation}
One can readily solve the corresponding Wheeler-DeWitt equation
by again introducing the variables $\tau$ and $\xi$,
Eqs.\ (\ref{tau}) and (\ref{xi}). Since
$\pi_{\phi}$ is conserved, we obtain the basis solutions
\begin{equation}
\Phi_{\omega p}(\tau,\xi,\phi) = \frac{1}{\sqrt{2\eta(\omega,p)}}
f_{\eta(\omega,p)}(\tau)S_{\omega}(\xi)\frac{e^{ip\phi}}{\sqrt{2\pi}}\,,
\end{equation}
where $\eta(\omega,p) \equiv \sqrt{\omega^2 + p^2}$. The corresponding $L^2$
basis functions, $\Psi_{\omega p}$, are given by
\begin{equation}
\Psi_{\omega p}(\tau,\xi,\phi)
= \frac{f_{\eta(\omega,p)}(\tau)}{\left| f_{\eta(\omega,p)}(\tau)\right|}
S_{\omega}(\xi)\frac{e^{ip\phi}}{\sqrt{2\pi}}\,.
\end{equation}
and the effective Hamiltonian, $H_{\it eff}$, is
\begin{equation}
H_{\it eff}\Psi_{\omega p}(\tau,\xi,\phi) =
\frac{\eta(\omega,p)}{\left|f_{\eta(\omega,p)}(\tau)\right|^2}
\Psi_{\omega p}(\tau,\xi,\phi)\,. \label{hefts}
\end{equation}

One advantage of this model is that there is more than one dynamical
variable present, so that one could regard one of the variables
-- say, $\phi$, since classically it evolves monotonically --
to be a ``physical clock", against which
the evolution of the other variable, $\xi$, is to be measured.
However, since $f_{\eta(\omega,p)}(\tau) \to \infty$
as $\tau \to \infty$, it is clear that
all of the dynamical variables, including the ``physical clock", will
``freeze" in the classically forbidden region. Thus, we see that the
``stoppage of motion" predicted in this model actually
would be physically unobservable; rather, this freezing
would correspond, physically, to measurable time stopping
at the maximum expansion of the universe.

It would be of interest to investigate whether the above
``freezing'' phenomenon persists for other choices of time
slicing -- particularly the $\alpha$-slicing -- and, if so,
which variables ``freeze''.  Our calculations above for the
$\tau$-slicing were enormously simplified by the presence of an
orthogonal basis of
solutions in a separated form. Since the
$\tau$-derivatives of these basis solutions are proportional to themselves,
$C_t$ is diagonal in this basis. Thus, one can easily obtain
$C_t^{\dagger}$, and, thereby, calculate $H_{\it eff}$. On the other hand,
for the $\alpha$-slicing, although it is not difficult to calculate $C_t$
directly from the definition
(\ref{Ct}), we do not thereby obtain simple expression for $C_t^{\dagger}$.
For this reason, we have
been unable to calculate $H_{\it eff}$ and study the issue of ``freezing'' in
the $\alpha$-slicing.
\end{section}
\setcounter{equation}{0}

\begin{section}*{Acknowledgments}
We thank A. Ashtekar, A.O. Barvinsky, B.K. Berger,
R. Geroch, P. H\'{a}j\'{\i}\v{c}ek, A. Hosoya, B.S. Kay,
H. Kodama, K. Kucha\v{r}, J. Louko, C. Rovelli
and R.S. Tate for useful discussions.
This work was supported in part by Schweizerischer Nationalfonds
and the US
National Science Foundation
under Grant No. PHY 92-20644.
\end{section}

\renewcommand{\theequation}{A\arabic{equation}}
\begin{section}*{Appendix A:
An alternative quantization of class A Bianchi models}

In Sec.\ \ref{Hc} we eliminated the momentum constraint of
the class A Bianchi models by noting that, classically, we can
choose $h_{\alpha\beta}$ and $\pi^{\alpha\beta}$ to be diagonal. We then
wrote down the Wheeler-DeWitt equation for the diagonal Bianchi
metrics and found it to have the form of a
Klein-Gordon equation in a 3-dimensional flat spacetime with
a time-dependent potential.  However, it is more natural not
to impose the diagonal form of the metric at the outset, and to
proceed by treating
the Hamilitonian and momentum constraints on an equal footing.
We employ this latter approach in this Appendix. For simplicity,
we specialize to the vacuum models and will focus upon the
Bianchi I and Bianchi IX cases. Many of the results here,
including the algebra of the momentum constraints, have been obtained by
Ashtekar and Samuel \cite{AS}.

We consider, first, the form of the Wheeler-DeWitt equation for
the Bianchi models when the restriction to diagonal metrics is not imposed.
For each Bianchi Lie group, $G$, mini\-superspace is
comprised of the 6-dimensional
manifold of left invariant metrics, $h_{ab}$, on $G$.
The Hamiltonian constraint is
\begin{equation}
H = G_{abcd}\pi^{ab}\pi^{cd} + V = 0\,,  \label{supham}
\end{equation}
where the inverse supermetric is
\begin{equation}
G_{abcd} = \frac{1}{2}h^{-1/2}\left( h_{ac}h_{bd} + h_{ad}h_{bc}-h_{ab}h_{cd}
\right)
\end{equation}
and the determinant, $h$, of $h_{ab}$ is defined with respect to a fixed,
left invariant volume element on $G$.
It should be noted that the two lower
index pairs $(ab)$ and $(cd)$ correspond, respectively, to the upper
indices $A$ and $B$ in Eq.\ (\ref{2suha}) of Sec.\ \ref{Prop}.
Thus, the supermetric, $G_{AB}$, is
\begin{equation}
G^{abcd} =
\frac{1}{2}h^{1/2}\left( h^{ac}h^{bd} + h^{ad}h^{bc}- 2h^{ab}h^{cd}\right)\,.
\label{asup}
\end{equation}

The geometry of $G_{AB}$ was studied in the classic work of DeWitt
\cite{DW}. The signature of $G_{AB}$ is Lorentzian $(-+++++)$, and the
transformations $h_{ab} \to e^{2\alpha}h_{ab}$
define timelike, hypersurface orthogonal,
conformal isometries of $G_{AB}$. We may identify each of these
orthogonal hypersurfaces with the 5-dimensional manifold,
${\cal M}_C$, of left invariant
conformal metrics $\tilde{h}_{ab} = h^{-1/3}h_{ab}$ on $G$. The
supermetric then takes the form,
\begin{equation}
G_{AB} = e^{3\alpha}\left( -24 \nabla_A\alpha \nabla_B\alpha
+ H_{AB}\right)\,.
\end{equation}
The Riemannian metric, $H_{AB}$, on ${\cal M}_C$ is
invariant under the $SL(3,\reell)$ transformations
\begin{equation}
h_{ab} \to N_{a}^{\ c}N_{b}^{\ d}h_{cd}\,,
\end{equation}
where $N_{a}^{\ c}$ satisfies ${\rm det} N = 1$.
If we introduce arbitrary local coordinates
$\zeta^{\Lambda}$ (${\Lambda}=1-5$) on
${\cal M}_C$, then the coordinate components of $H_{AB}$ take the
explicit form
\begin{equation}
H_{{\Lambda} \Sigma} = \tilde{h}^{ab}
\frac{\partial\tilde{h}_{bc}}{\partial\zeta^{\Lambda}}\tilde{h}^{cd}
\frac{\partial\tilde{h}_{da}}{\partial\zeta^\Sigma} \label{gab}\,,
\end{equation}
where $\tilde{h}^{ab}$ is the inverse of $\tilde{h}_{ab}$.
The Ricci tensor and the scalar curvature of $H_{AB}$ are
\begin{eqnarray}
\bar{R}_{AB} & = & -\frac{3}{4} H_{AB}\,, \label{ricci} \\
\bar{R} & = & -\frac{15}{4} \label{scal}
\end{eqnarray}
(see Ref.\ \cite{DW}, Appendix A, for details; we have corrected the
corresponding equations of that reference by a
factor of $1/2$.) Thus, in contrast to the diagonal case,
the geometry of mini\-superspace is now curved.
Thus, the Wheeler-DeWitt equation (\ref{kgeq}) corresponding to
(\ref{supham}) now takes the form of a curved space wave equation
\begin{equation}
\left[ \frac{\partial^2\ }{\partial\alpha^2}
- 24 D_A D^A - 90\xi + 24V \right]\Phi = 0\,,  \label{conc}
\end{equation}
where $D_A$ is the derivative operator associated with $H_{AB}$.

For the Bianchi I model, no momentum constraints are present
so the full quantum constraints will be imposed on $\Phi$ by
satisfying Eq.\ (\ref{conc}) with $V=0$. Since Eq.\ (\ref{conc}) is
a curved space equation, it is easily seen that the theory obtained
in this manner is not, in any sense, equivalent to the theory obtained by
restricting to diagonal metrics as done in Sec.\ \ref{Hc}.
(The reasons for this inequivalence were explained in that section.)

One further feature of Eq.\ (\ref{conc}) in the Bianchi I
model is worth noting. Since
mini\-superspace is six dimensional, the natural \cite{HK},
conformally invariant
version of the Wheeler-DeWitt equation
is obtained by choosing $\xi = 1/5$.  However,
the lower bound of the spectrum of $-D_A D^A$ is
$1/2$ (see, e.g., Ref.\ \cite{T}, p.~49).  Thus, the Wheeler-DeWitt
equation with this choice of $\xi$ is ``tachyonic", and there exist
spatially well behaved solutions which grow exponentially
as $\alpha \to \pm\infty$.  In particular, the
conformally invariant choice of Wheeler-Dewitt equation does not
allow the division of the space of solutions into
asymptotic positive and negative frequency
subspaces as $\alpha \to -\infty$,
i.e., our prescription for the construction of ${\cal H}$ breaks down.
However, this difficulty can be avoided by making a
different choice of $\xi$.

The classical momentum constraints
\begin{equation}
-2 D_a \pi^{ab} = 0
\end{equation}
are nontrivial in all the other Bianchi models. (Here $D_a$ denotes the
derivative operator on $G$
associated with the left invariant metric $h_{ab}$.)
In order to motivate a
natural operator version of these constraints, we rewrite them as follows.
First, if $s^a$ is a left invariant vector field on $G$, then by a direct
computation using
the formula for the connection
in terms of the structure constant field $C^{c}_{\ ab}$
(see, e.g., Ref.\ \cite{MTW}, p.\ 314), we obtain
$D_{a} s^a = C^{b}_{\ ba}s^{a} = A_a s^a = 0$.
Hence, we find that the momentum constraint
\begin{equation}
P_{s} \equiv -2 s^c h_{bc} D_a \pi^{ab} \label{ps}
\end{equation}
can be rewritten as
\begin{eqnarray}
P_{s} & = & -2 D_a (s^c h_{bc} \pi^{ab})
+ 2 (D_{a} s_{b}) \pi^{ab} \nonumber \\
& = & ({\cal L}_s h_{ab}) \pi^{ab}\,,
\label{ps2}
\end{eqnarray}
since the first term on the right side of the first line of this equation
is the divergence of a left invariant vector field and, thus, vanishes.

When we promote $P_{s}$ to an operator, it is natural to choose the
factor ordering (\ref{ps2}), since then the constraint $P_s \Phi= 0$
implies that the wavefunction $\Phi$ is
invariant under the transformation $\delta h_{ab} = {\cal L}_{s}h_{ab}$.
(As is well known, an analogous choice and interpretation of the
momentum constraint operators can be made in full quantum gravity
\cite{DW}, the only difference being that neither $h_{ab}$ nor $s^a$
are restricted to be left invariant.)
Note that with this choice of $P_{s}$, we have
\begin{eqnarray}
\left[ h_{ab}, P_s\right] & = & i{\cal L}_s h_{ab}, \\
\left[ \pi^{ab}, P_s\right] & = & i{\cal L}_s \pi^{ab}\,,
\end{eqnarray}
and
\begin{equation}
\left[ P_s, P_t\right] = iP_{\left[ s,t \right]}\,.
\end{equation}

Let us study the imposition of the momentum constraints for the
case of the Bianchi IX model in some detail. The Lie group, $G$, of
the Bianchi IX model is $SU(2)$, and the
inverse, $(M^{-1})_{ab}$,
of the left invariant tensor field
$M^{ab}$ defined by Eq.\ (\ref{eps}) provides a positive definite metric
on the 3-dimensional vector space, $W$,
of left invariant vector fields. The exponentiated version of the
momentum constraints $P_s \Phi=0$ imply that the allowed
wavefunctions are invariant under the transformation
$$
{h}_{ab} \to O_{a}^{\ c}O_{b}^{\ d} {h}_{cd} \,,
$$
where $O_{a}^{\ c}$ is an arbitrary
orthogonal linear map [with respect
to $(M^{-1})_{ab}$]
on $W$ with positive determinant.

It is convenient to
introduce the following parametrization of mini\-superspace. Since $h_{ab}$
defines a positive
self-adjoint map on $W$ in the inner product $(M^{-1})_{ab}$, we may
characterize each $h_{ab}$ (i.e., point in mini\-superspace)
by its eigenvalues,
$\lambda_1 \geq \lambda_2 \geq \lambda_3 > 0$, and the rotation,
$O$, which carries its triad of eigenvectors into some fixed, orthonormal
basis.
Note, however, that this parametrization becomes singular when degeneracy
occurs in the eigenvalues of $h_{ab}$.

The Hilbert space of states, ${\cal H}$, is constructed from functions
$\Phi (\lambda_1, \lambda_2, \lambda_3, O)$ on
mini\-superspace which satisfy the momentum and Hamiltonian
constraints. The momentum constraints are easily imposed,
since they simply require $\Phi$ to be independent of
$O$. The form of the Wheeler-DeWitt
equation in these coordinates can be obtained as follows.
On the manifold, ${\cal M}_C$, of conformal
metrics (where $\lambda_1\lambda_2\lambda_3 = 1$, so only two of
the $\lambda_i$'s are independent), we have from Eq.\ (\ref{gab})
\begin{eqnarray}
H_{\Lambda \Sigma} & = &
{\rm Tr}\left[ \tilde{h}^{-1}
\frac{\partial\tilde{h}}{\partial\zeta^{\Lambda}}\tilde{h}^{-1}
\frac{\partial\tilde{h}}{\partial\zeta^{\Sigma}}\right]
\nonumber \\
 & =  & -{\rm Tr}\left(\frac{\partial T}{\partial\zeta^{\Lambda}}
\frac{\partial T^{-1}}{\partial\zeta^{\Sigma}}\right)
- 2{\rm Tr}\left[ O^{-1}\frac{\partial O}{\partial\zeta^{\Lambda}}
O^{-1}\frac{\partial O}{\partial\zeta^{\Sigma}} -
O^{-1}\frac{\partial O}{\partial\zeta^{\Lambda}} T
O^{-1}\frac{\partial O}{\partial\zeta^{\Sigma}} T^{-1}\right]\, . \label{pp}
\end{eqnarray}
where $T_{ab}$ is the diagonal matrix with eigenvalues
$\lambda_1, \lambda_2, \lambda_3 $.
By writing
\begin{equation}
O^{-1}dO \equiv \left( \begin{array}{ccc}
0 & -X_3 & X_2 \\ X_3 & 0 & -X_1 \\ -X_2 & X_1 & 0 \end{array}
\right) \,,
\end{equation}
we obtain \cite{T}
\begin{eqnarray}
H_{\Lambda \Sigma}d\zeta^{\Lambda} d\zeta^{\Sigma}
& =  & (d\ln\lambda_1)^2 + (d\ln\lambda_2)^2
+ (d\ln\lambda_3)^2 \nonumber \\
 & & + \lambda_1 (\lambda_2 -\lambda_3)^2 X_1^2
+ \lambda_2 (\lambda_3 -\lambda_1)^2 X_2^2
+ \lambda_3(\lambda_1-\lambda_2)^2 X_3^2 \,.
\end{eqnarray}

Although the momentum constraints require
$\Phi$ not to depend upon $O$, the portion of the
supermetric involving $(X_1,X_2,X_3)$ makes a $\lambda$-dependent
contribution to the volume element on superspace, and this, in turn,
nontrivially affects the form of Wheeler-DeWitt equation. Taking into
account the condition $\lambda_1\lambda_2\lambda_3 = 1$ on
${\cal M}_C$, we see that this portion of the metric
contributes to $\sqrt{H}$ the additional factor
\begin{equation}
C = \left| \prod_{i>j}(\lambda_i - \lambda_j)\right| \,.
\end{equation}
Writing $\lambda_1 \equiv e^{2(\beta_{+}+\sqrt{3}\beta_{-})}$,
$\lambda_2 \equiv e^{2(\beta_{+}-\sqrt{3}\beta_{-})}$, and
$\lambda_3 \equiv e^{-4\beta_{+}}$, we obtain
\begin{equation}
  C(\beta_{\pm})
  = 8 \left| \sinh2\sqrt{3}\beta_{-}
\sinh(3\beta_{+}-\sqrt{3}\beta_{-})
\sinh(3\beta_{+}+\sqrt{3}\beta_{-})\right| \,.
\end{equation}
The Wheeler-DeWitt equation then becomes
\begin{equation}
\left[ \frac{\partial^2\ }{\partial\alpha^2} -
\frac{1}{C(\beta_{\pm})}
\sum_{i=\pm} \frac{\partial\ \ }{\partial\beta_{i}}
C(\beta_{\pm})\frac{\partial\ \ }{\partial\beta_i}
+ 24V - 90\xi\right]\Phi  =   0\,.
\end{equation}
The variables $\beta_{\pm}$ are restricted by the condition
$\lambda_1 \geq \lambda_2 \geq \lambda_3 > 0$, but we may
drop this restriction by requiring, instead, the
invariance of $\Phi$ under the transformations of
$\beta_{\pm}$ corresponding to permutations of $\lambda_i$ $(i = 1,2,3)$.
The kinetic term in the Wheeler-DeWitt equation
can be converted into flat spacetime wave operator by rescaling $\Phi$ as
$\Phi \equiv C(\beta_{\pm})^{-1/2}\tilde{\Phi}$.
Then, we obtain
\begin{equation}
\left[ \frac{\partial^2\ }{\partial\alpha^2} -
\frac{\partial^{2}\ \ }{\partial\beta_{+}^2}
-\frac{\partial^{2}\ \ }{\partial\beta_{-}^2}
+ 24\tilde{V} -90\xi \right]\tilde{\Phi}  =  0
\end{equation}
with
\begin{equation}
\tilde{V} \equiv V + \frac{1}{24C(\beta_{\pm})^{1/2}}\left[
\frac{\partial^2\ \ }{\partial\beta_{+}^2}
+\frac{\partial^2\ \ }{\partial\beta_{-}^2}\right]C(\beta_{\pm})^{1/2}\,.
\end{equation}
This differs substantially from the form of the Wheeler-DeWitt equation
derived in Sec.\ \ref{Hc}.
\end{section}

\newpage
\begin{section}*{Figure captions}
Figure 1: $d\phi/d\alpha \equiv \partial H_{\it eff}/\partial p$
for $p = 4,16,64,256$.  The dotted line is the corresponding classical value
 $[1- e^{4\alpha}/p^2]^{-1/2}$.

\

\noindent
Figure 2: The region swept by the interval of $\phi$
(for given $\alpha$) containing
90\% of the squared $L^2$ wavefunction for a Gaussian wavepacket with
$\langle p \rangle = 64$ and
$\Delta  p \equiv [\langle p^2 \rangle - \langle p \rangle^2]^{1/2} = 10$
in the Robertson-Walker model with a massless scalar field.  The solid
line is the corresponding classical solution.

\

\noindent
Figure 3:
The region swept by the interval of $\bar{\xi}$
containing 90\% of the
squred $L^2$ wavefunction for a Gaussian wavepacket with
$\langle \bar{\omega}\rangle = 32$ and $\Delta \bar{\omega} = 5$ in the
Taub model, where we have
defined $\bar{\omega} \equiv 3^{-1/2}\omega$,
$\bar{\xi} \equiv 3^{1/2}\xi$,
$\bar{\tau} \equiv 3^{1/2}\tau$. The solid line is the corresponding
classical solution.
\end{section}

\begin{thebibliography}{99}
\bibitem{COM1} The Hamiltonian does not vanish for
asymptotically-flat spacetimes, but
the same problems arise when one
evolves the system with respect to slices which do not move at spacelike
infinity.

\bibitem{W} R.M. Wald, Phys. Rev. D {\bf 48}, R2377 (1993).

\bibitem{KAR} K. Kucha\v{r}, J. Math. Phys. {\bf 22}, 2640 (1981).

\bibitem{MH} C. Misner, in {\it Magic without Magic: John Archibald Wheeler}
edited by J. Klauder (Freeman, San Francisco, 1972).  See also J.J. Halliwell,
Phys. Rev. D {\bf 38}, 2468 (1988).

\bibitem{HK} P. H\'aj\'{\i}\v{c}ek and K.V. Kucha\v{r}, Phys. Rev. D
{\bf 41}, 1091 (1990).

\bibitem{AMK} A. Ashtekar and A. Magnon, Proc. R. Soc. London {\bf A346},
375 (1975); B.S. Kay, Commun. Math. Phys. {\bf 62}, 55 (1978).

\bibitem{ash} A. Ashtekar {\it Lectures on Non-Perturbative
Canonical Gravity} (World Scientific, Singapore, 1991).

\bibitem{ADM} R. Arnowitt, S. Deser, and C.W. Misner, in {\it Gravitation:
An Introduction to Current Research}, edited by L. Witten (Wiley, New York,
1962).

\bibitem{KUCH} K. Kucha\v{r}, {\it Proc. 4th Canadian Conf. General Relativity
and Relativistic Astrophysics}, edited by G. Kunstatter, D. Vincent,
and J. Williams (World Scientific, Singapore, 1992).

\bibitem{BAR} A. O. Barvinsky, Class. Quantum Grav. {\bf 10}, 1957 (1993).

\bibitem{HIG} A. Higuchi, in preparation.

\bibitem{COM0} Inclusion of a scalar field turns out to be essential in
some models. We only consider the case with one scalar field for simplicity,
but the number of scalar fields does not affect our discussion.

\bibitem{MCLM} M.A.H. MacCallum, in {\it General Relativity: An Einstein
Centenary Survey}, edited by S.W. Hawking and W. Israel
(Cambridge University, Cambridge, 1979).

\bibitem{WTEX} R.M. Wald, {\it General Relativity} (University of Chicago
Press,
Chicago, 1984).

\bibitem{E} G.F.R. Ellis and M.A.H. MacCallum, Commun. Math. Phys. {\bf 12},
108 (1969).

\bibitem{MAC} M.A.H. MacCallum and A.H. Taub, Commun. Math. Phys. {\bf 25},
173 (1972); M.P. Ryan, J. Math. Phys. {\bf 15}, 812 (1972); G.E. Sneddon,
J. Phys. A: Math. Gen. {\bf 9}, 229 (1976).

\bibitem{HOSO} Compact spaces admitting ``locally homogeneous'' actions of
Bianchi groups have been analyzed recently by
T. Koike, M. Tanimoto and A. Hosoya,
{\it Compact homogeneous universes},
Tokyo Institute of Technology preprint TIT/HEP-208/COSMO, gr-qc/9405052,
J. Math. Phys., in press.

\bibitem{W2} R.M. Wald, Phys. Rev. D {\bf 28}, R2118 (1983).

\bibitem{KS} R. Kantowski and R.K. Sachs, J. Math. Phys. {\bf 7}, 443 (1966).

\bibitem{RTS} For a related observation, see J.D. Romano and R.S. Tate,
Class. Quantum Grav. {\bf 6}, 1487 (1989); K. Schleich, {\it ibid}. {\bf 7},
1529 (1990).

\bibitem{CH} P. Chmielowski,
Class. Quantum Grav. {\bf 11}, 41 (1994).

\bibitem{KW} B.S. Kay and R.M. Wald, Phys. Rep. {\bf 207},
49 (1991).

\bibitem{WBOOK}  R.M. Wald {\it Quantum Field Theory in Curved
Spacetime and Black Hole Thermodynamics} (University of Chicago Press,
Chicago, 1994).

\bibitem{COM5} The validity of this approximation has not been rigorously
proven.  For a numerical evidence in the classical dynamics, see
B.K. Berger, Gen. Relativ. Gravit. {\bf 23}, 1385 (1991); Phys. Rev. D
{\bf 49}, 1120 (1994).

\bibitem{MIX} C.W. Misner, Phys. Rev. Lett. {\bf 22}, 1071 (1969).

\bibitem{HI} For other examples of
nonchaotic behavior of Bianchi VIII and XI
models with additional degrees of freedom, see P. Halpern, Gen. Relativ.
Gravit. {\bf 19}, 73 (1987); H. Ishihara, Prog. Theor. Phys. {\bf 74}, 1490
(1985).

\bibitem{WALL} S.M. Waller, Phys. Rev. D {\bf 29}, 176 (1984).

\bibitem{GR} I.S. Gradshteyn and I.M. Ryzhik, {\it Table of Integrals,
Series, and Products} (Academic Press, New York, 1980).

\bibitem{CK} The WKB approximation for this equation (for the solutions
which decay for $\alpha \to +\infty$) can be found in
C. Kiefer, Phys. Rev. D {\bf 38}, 1761 (1988).

\bibitem{S} L.I. Schiff, {\it Quantum Mechanics} (McGraw-Hill, New York,
1955), p.~271.

\bibitem{COM4} Wavepackets spread out in general as in the usual quantum
mechanics \cite{CK}. However, even this spreading does not occur in this
model due to the fact that it is approximately the 2-dimensional massless
scalar field for large and negative $\alpha$.

\bibitem{HH} J.B. Hartle and S.W. Hawking, Phys. Rev. D {\bf 28}, 2960
(1983).

\bibitem{H} S.W. Hawking, Nucl. Phys. {\bf B239}, 257 (1984).

\bibitem{V1} A. Vilenkin, Phys. Rev. D {\bf 33}, 3560 (1986).

\bibitem{V2} A. Vilenkin, Phys. Rev. D {\bf 37}, 888 (1988).

\bibitem{V3} A. Vilenkin, ``Approaches to Quantum Cosmology", to be
published.

\bibitem{ATU} For the quantization of this model via a
different approach, see
A. Ashtekar, R. Tate and C. Uggla,
Int. J. Mod. Phys. {\bf D2}, 15 (1993).

\bibitem{MR} S. Martinez and M. Ryan, in {\it Relativity, Cosmology,
Topological Mass and Supergravity}, Proceedings of the Fourth Silarg
Symposium, Caracas, Venezuela, 1982, edited by C. Aragone (World Scientific,
Singapore, 1983).

\bibitem{HMS} A. Higuchi, G.E.A. Matsas, and D. Sudarsky, Phys. Rev. D
{\bf 46}, 3450 (1992).

\bibitem{AS} A. Ashtekar and J. Samuel, Class. Quantum Grav. {\bf 8}, 2191
(1991).

\bibitem{DW} B.S. DeWitt, Phys. Rev. {\bf 160}, 1113 (1967).

\bibitem{T} A. Terras, {\it Harmonic Analysis on Symmetric Spaces and
Applications II} (Springer, Berlin, 1988).

\bibitem{MTW} C.W. Misner, K.S. Thorne, and J.A. Wheeler, {\it Gravitation}
(Freeman, San Francisco, 1972).
\end{thebibliography}
\end{document}